\documentclass[prd,showpacs,preprintnumbers,amsmath,amssymb,floatfix]{revtex4}
\usepackage{graphicx}
\usepackage{dcolumn}
\usepackage{bm}
\usepackage{amssymb}
\usepackage{epsfig}
\usepackage{color}

\newcommand{\ba}{\begin{eqnarray}}
\newcommand{\ea}{\end{eqnarray}}
\newcommand{\be}{\begin{equation}}
\newcommand{\ee}{\end{equation}}
\newcommand{\bdisplay}{\begin{displaymath}}
\newcommand{\edisplay}{\end{displaymath}}

\newcommand{\eq}[1]{Eq.\,(\ref{#1})}

\newcommand{\calF}{\mbox{${\cal F}$}}

\begin{document}

\title{Applications of the  leading-order  Dokshitzer-Gribov-Lipatov-Altarelli-Parisi evolution equations to the combined HERA data on deep inelastic scattering}
\author{Martin~M.~Block}
\email{mblock@northwestern.edu}
\affiliation{Department of Physics and Astronomy, Northwestern University,
Evanston, IL 60208, USA}
\author{Loyal Durand}
\email{ldurand@hep.wisc.edu}
\altaffiliation{Mailing address: 415 Pearl Ct., Aspen, CO 81611}
\affiliation{Department of Physics, University of Wisconsin, Madison, WI 53706, USA}
\author{Phuoc Ha}
\email{pdha@towson.edu}
\affiliation{Department of Physics, Astronomy and Geosciences, Towson University, Towson, MD 21252, USA}
\author{Douglas W. McKay}
\email{dmckay@ku.edu}
\affiliation{Department of Physics and Astronomy, University of Kansas, Lawrence, KS 66045, USA}
\date{\today}

\begin{abstract}
We recently derived explicit solutions of the leading-order   Dokshitzer-Gribov-Lipatov-Altarelli-Parisi (DGLAP) equations for the $Q^2$ evolution of the singlet structure function $F_s(x,Q^2)$ and  the gluon distribution $G(x,Q^2)$ using very efficient Laplace transform techniques.  We apply our results here to a study of the HERA data on deep inelastic $ep$ scattering as recently combined by the H1 and ZEUS groups.  We use initial distributions $F_2^{\gamma p}(x,Q_0^2)$ and $G(x,Q_0^2)$ determined for $x<0.1$ by a global fit to the HERA data, and extended to $x=1$ using the shapes of those distributions determined in the CTEQ6L and MSTW2008LO analyses from fits to other data. Our final results are insensitive at small $x$ to the details of the extension. We  obtain the singlet quark distribution $F_s(x,Q_0^2)$ from $F_2^{\gamma p}(x,Q_0^2)$ using small nonsinglet quark distributions taken from either the CTEQ6L or the MSTW2008LO analyses, evolve $F_s$ and $G$ to arbitrary $Q^2$, and then convert the results to individual quark distributions.  Finally, we show  directly from a study of systematic trends in a comparison of the evolved $F_2^{\gamma p}(x,Q^2)$ with the HERA data, that the assumption of leading-order DGLAP evolution is inconsistent with those data.

\end{abstract}


\maketitle


\section{Introduction}\label{sec:Introduction}

In recent papers \cite{bdhmLO2,bdhmNLO}, we showed that it is possible to  solve the coupled leading-order (LO)  Dokshitzer-Gribov-Lipatov-Altarelli-Parisi (DGLAP) evolution equations \cite{dglap1,dglap2,dglap3}  for the singlet quark structure function $F_s(x,Q^2)=\sum_ix[q_i(x,Q^2)+\bar{q}_i(x,Q^2)]$ and the gluon distribution $G(x,Q^2)=xg(x,Q^2)$ directly using a method based on Laplace transforms.  While the method is formally equivalent through the known connection between Laplace and Mellin transforms \cite{HTF4} to methods based on the latter -- see, {\em e.g.} \cite{dglap1,Mellin}, we find the present approach to be clearer intuitively and much more efficient numerically.  In particular,  the distributions $F_s(x,Q^2)$ and $G(x,Q^2)$  at  a virtuality $Q^2$ can be expressed simply as  convolutions of the distributions $F_s(x,Q_0^2)$ and $G(x,Q_0^2)$ at a starting value $Q_0^2$  with analytically defined kernels in the ordinary variables.  Alternatively, the results can  be expressed as inverse Laplace transforms of products of the kernels in Laplace space with the Laplace transforms of the initial distributions.

We perform the inverse Laplace transforms necessary in our approach using very fast and accurate new numerical  algorithms  \cite{inverseLaplace1,inverseLaplace2}. These do not require that we work on a preassigned numerical grid, and make the solution of the evolution equations at arbitrary values $x$ and $Q^2$ straightforward on desktop or laptop computers.  We have extended the Laplace method  elsewhere \cite{bdhmNLO} to next-to-leading order (NLO) in $\alpha_s$, including to nonsinglet distributions,  but will not pursue that extension  here.

In the present paper, we apply these methods to test the consistency of the assumed LO  evolution of the structure functions with the HERA data \cite{ZEUS1,ZEUS2,H1} on deep inelastic $ep$ (or $\gamma^*p$) scattering, using those data as recently combined by the H1 and ZEUS experimental groups \cite{HERAcombined}. As shown earlier \cite{bdm1,bdm2}, if a LO treatment of the DGLAP evolution is sufficient, the necessary starting distribution $G_0(x)\equiv G(x,Q_0^2)$  can be obtained from a global fit to the structure function $F_2^{\gamma p}(x,Q^2)$ by requiring that the LO evolution equation for $F_2^{\gamma p}(x,Q^2)$ be satisfied for $Q^2=Q_0^2$. Both $F_2^{\gamma p}(x,Q_0^2)$ and $G(x,Q_0^2)$ are then determined  directly by experiment.

To obtain our starting distributions, we perform the required global fit to $F_2^{\gamma p}$ using the HERA data  for $x<0.1$, and extend the fit to $x=1$ using the shape of that distribution as determined in the CTEQ6L \cite{CTEQ6L} and MSTW2008LO  \cite{MSTW1} analyses which included other DIS data at large $x$. Our final results at small $x$ are insensitive to the details of the extension.  We pick as a starting value  for the $Q^2$ evolution a value $Q_0^2=4. 5$ GeV$^2$, which is well within the region of dense data, and determine the starting $G$ as described above.

The singlet distribution $F_s(x,Q^2)$ differs from $F_2^{\gamma p}(x,Q^2)$ by small nonsinglet contributions that depend primarily on the valence quark distributions, which agree fairly well for different LO analyses at moderate $Q^2$ (compare, {\em e.g.} \cite{CTEQ6L} and \cite{MSTW1}).  We will therefore  simply use the results of the CTEQ6L and MSTW2008LO analyses \cite{CTEQ6L,MSTW1} to make the necessary conversion from $F_2^{\gamma p}$ to $F_s$ at $Q_0^2=4.5$ GeV$^2$, and the evolved nonsinglet contributions to convert the evolved $F_s(x,Q^2)$  back to the function $F_2^{\gamma p}(x,Q^2)$ which can be compared to the HERA data for $Q^2\not=Q_0^2$.

We also combine the evolved $F_s$ with the nonsinglet distributions of CTEQ6L and MSTW2008LO to obtain a new set of CTEQ6L-like or MSTW-like quark distributions. Even though we use the same nonsinglet distributions as those authors, our final results differ from the originals because of  our use of the combined HERA data rather than the original H1 and ZEUS results, and, importantly, our use of starting distributions $F_s(x,Q_0^2)$ and $G(x,Q_0^2)$ determined directly from experiment up to the small nonsinglet contributions to the former.

We find that the evolved $F_2^{\gamma p}(x,Q^2)$ calculated  using LO DGLAP evolution differs {\em systematically} in its dependence on $x$ and $Q^2$ from the combined HERA data at values of $Q^2$ away from $Q_0^2$.  We conclude  that  LO DGLAP evolution is not consistent with the data, a conclusion reached less directly by other authors, {\em e.g.}, in \cite{MSTW1,CTEQ6.1,HERAcombined}.  We emphasize in this connection that the only fitting involved in our approach is  in the QCD-independent global fit to the data on $F_2^{\gamma p}$; we do not need to solve the complete set of evolution equations and then attempt to fit the data using the many input parameters typically introduced in the parameterization of initial parton distributions.

Our conclusion on the inconsistency of LO evolution is not surprising. Next-to-leading-order (NLO) effects on the evolution are known to be large. However, our results give a {\em  direct} demonstration of the necessity of going beyond LO  independent of the substantial complications that a NLO analysis entails.

In the Appendix, we present an accurate alternative method of testing LO evolution based on the exact LO evolution equation for $F_2^{\gamma p}(x,Q^2)$, and an approximate evolution equation for $G(x,Q^2)$.  Its advantage is that the input necessary to test the assumption of LO evolution can be obtained directly from the measured $F_2^{\gamma p}(x,Q^2)$.  The application of this method to the HERA data leads to the same conclusion as stated above: the assumption of  LO evolution is  inconsistent with HERA data.


\section{Preliminaries}\label{sec:preliminaries}

\subsection{Solution of the coupled evolution equations for $F_s$ and $G$}\label{subsec:solutions}

In the present paper, we use the method developed in detail in \cite{bdhmLO2,bdhmNLO} to solve the coupled DGLAP evolution equations for $F_s$ and $G$.  We will not give the details here, but note that our method is based on Laplace transforms. We first rewrite the evolution equations in terms of the variables  $v=\ln{(1/x)}$ and $Q^2$ instead of $x$ and $Q^2$. The integral coupling terms in the equations then  reduce to a form that involves convolutions in $v$, and the equations can be converted by Laplace transformation to factored  homogeneous first-order differential equations in $Q^2$ and a Laplace variable $s$, and solved directly.

Using the notation $\hat{F}_s(v,Q^2)\equiv F_s(e^{-v},Q^2)$,  $\hat{G}(v,Q^2)\equiv G(e^{-v},Q^2)$ for the distributions written in terms of $v$ and $Q^2$, and introducing the Laplace transforms
\ba
f_s(s,Q^2)&\equiv &{\cal L}\left[ \hat F_s(v,Q^2);s\right],\qquad
g(s,Q^2)\equiv {\cal L}[\hat G(v,Q^2);s],
\ea
we find that the Laplace-space distributions generated  by evolution from $Q_0^2$ to $Q^2$ can be expressed in terms of the initial distributions $f_{s0}(s)\equiv f_s(s,Q_0^2)$ and $g_0(s)\equiv g(s,Q_0^2)$ as
\ba
\label{f(s,tau)}
f_s(s,Q^2) &= &k_{ff}(s,\tau)f_{s0}(s)+k_{fg}(s,\tau) g_0(s), \label{f} \\
\label{g(s,tau)}
g(s,Q^2)&= &k_{gf}(s,\tau) f_{s0}(s) +k_{gg}(s,\tau)g_0(s) \label{g}.
\ea
The kernels $k(s,\tau)$ in Eqs.\ (\ref{f}) and (\ref{g}) are  given explicitly in  \cite{bdhmLO2,bdhmNLO}. They depend on $Q^2$ and $Q_0^2$ only through the variable
\be
\tau(Q^2,Q_0^2) = \frac{1}{ 4 \pi}\int_{Q_0^2}^{Q^2} \alpha_s(Q'^2)\,d(\ln Q'^2), \label{tau}
\ee
which vanishes for $Q^2=Q_0^2$, with $k_{ff}(s,0)=k_{gg}(s,0)=1$ and $k_{fg}(s,0)=k_{gf}(s,0)=0$.
The kernels also depend on the number $n_f$ of active quarks.

If we have parametrized the initial distributions accurately analytically, and Laplace transformed the results to obtain $f_{s0}(s)$ and $g_0(s)$, we can calculate the inverse Laplace transforms of $f_s(s,Q^2)$ and $g(s,Q^2)$ directly to obtain the evolved distributions $\hat{F}_s(v,Q^2)$ and $\hat{G}(v,Q^2)$, with
\ba
\label{F_1}
\hat F_s(v,Q^2 )&=& {\cal L}^{-1}\left\{[k_{ff}(s,\tau)f_{s0}(s);v]+ [k_{fg}(s,\tau)g_0(s);v]\right\}, \\
\label{G_1}
\hat G(v,Q^2) &=&  {\cal L}^{-1}\left\{[k_{gf}(s,\tau)f_{s0}(s);v]+[k_{gg}(s,\tau)g_0(s);v]\right\}.
\ea

Alternatively, using the convolution theorem  to write the transforms of the products on the right-hand sides as convolutions, and using the fact that the inverse transforms of $f_{s0}(s)$ and $g_0(s)$ are the initial $v$-space distributions $\hat{F}_{s0}(v)=\hat{F}_{s}(v,Q_0^2)$, $\hat{G}_0(v)=\hat{G}(v,Q_0^2)$, we can write the solutions in the more intuitive form
\ba
\hat F_s(v,Q^2)&=&\int_0^v K_{FF}\left(v-w,\tau(Q^2,Q_0^2)\right)\hat F_{s0}(w)\,dw +\int_0^v K_{FG}\left(v-w,\tau(Q^2,Q_0^2)\right)\hat G_0(w)\,dw, \label{F} \\
\hat G(v,Q^2)&=& \int_0^v K_{GF}\left(v-w,\tau(Q^2,Q_0^2)\right)\hat F_{s0}(w)dw + \int_0^v K_{GG}\left(v-w,\tau(Q^2,Q_0^2)\right)\hat G_0(w)\,dw,  \label{G}
\ea
where the $v$-space  kernels $K_{FF},\ K_{FG},\ K_{GF}$ and $K_{GG}$, given by the inverse Laplace transforms of the corresponding $k's$,  describe the smearing and growth of the original distributions $\hat{F}_{s}(v,Q_0^2)$ and $\hat{G}(v,Q_0^2)$ through QCD radiation and splitting processes.

The inverse Laplace transforms needed to implement  Eqs.\ (\ref{F_1}) and (\ref{G_1})  can be calculated  efficiently using the very accurate and extremely fast algorithms  discussed in \cite{inverseLaplace1,inverseLaplace2}; these were used in the calculations reported here, and the results then converted to distributions in $x$ and $Q^2$.  The numerical techniques needed are discussed in detail in the Appendix in \cite{bdhmLO2}. These allow the fast solution of the complete set of DGLAP evolution equations on a standard desktop or laptop computer. The kernel technique will be discussed elsewhere.

The one-step inversion in Eqs.\ (\ref{F_1}) and (\ref{G_1})  is particularly useful in the case of devolution from large to small $Q^2$: the variable $\tau$ is then negative, the integrals that define $K_{FF}$ and $K_{GG}$ do not converge as ordinary integrals, and those kernels must be defined as generalized functions. This problem does not appear with the forms in Eqs.\ (\ref{F_1}) and (\ref{G_1}) provided $\hat{F}_s(v,Q^2)$ and $\hat{G}(v,Q^2)$ vanish sufficiently rapidly for $v\rightarrow 0$ that the products in Eqs.\ (\ref{f(s,tau)}) and (\ref{g(s,tau)}) vanish as a power of $1/s$ for $s\rightarrow\infty$. These conditions are satisfied in practice.

The evolved $\hat{F}_s(v,Q^2)$ and $\hat{G}(v,Q^2)$ must be continuous at quark thresholds where $n_f$ changes. We treat  the thresholds in $Q^2$ as in \cite{CTEQ6L,MSTW1,CTEQ6.1}.  In the course of the evolution from the initial $Q_0^2$ to a larger final virtuality, $Q^2$ may cross a threshold at $Q^2=M_i^2$ where quark $i$ becomes active, and the number  $n_f$ of active quarks increases by 1. This changes $n_f$-dependent coefficients in the evolution equations. However, the continuity of $\hat{F}_s(v,Q^2)$ and $\hat{G}(v,Q^2)$ as functions of $Q^2$ is guaranteed if we evolve first from $Q_0^2$ to $M_i^2$, take the results at $Q^2=M_i^2$ as new starting distributions, and then continue the evolution from $M_i^2$ to $Q^2$ with $n_f\rightarrow n_f+1$. We otherwise neglect mass effects on the evolution. The same remarks apply to the case of devolution from $Q_0^2$ to a smaller $Q^2$, with $n_f$ then decreasing by 1 at each transition.

We have checked that our methods accurately reproduce the LO results of CTEQ6L \cite{CTEQ6L}  for the evolution of $F_s$ and $G$ when we use starting distributions taken from their published results. The errors in the evolved distributions are $\lesssim 0.05$\% for CTEQ6L, as discussed in \cite{bdhmLO2}.  Similarly, we reproduce the results of MSTW2008LO   \cite{MSTW1} for the evolved $F_s$ and $G$ to $\lesssim 0.1-0.5$\%  \cite{bdhmLO2}.

The solution of the nonsinglet evolution equations for quark distributions such as $xq_i^-(x,Q^2)=x\left[q_i(x,Q^2)-\bar{q}_i(x,Q^2)\right]$ is simpler because of the absence of any coupling to the gluon distribution. The results in LO have the form \cite{bdhmNLO}
\be
\label{NS_form}
\hat{F}_{ns}(v,Q^2)={\cal L}^{-1}\left[k_{ns}(s,\tau)f_{ns,0}(s);v\right],
\ee
where $k_{ns}(s,\tau)$ is the common LO non singlet evolution kernel and $f_{ns,0}={\cal L}\left[\hat{F}_{ns}(v,Q_0^2);s\right]$.

We have discussed the  generalization of these results to next-to-leading order in \cite{bdhmNLO}. The decoupling of the evolution equations in that case requires a double Laplace transform and is considerably more complicated in detail, but can still be carried through analytically.  We will not pursue that here.


\subsection{Determination of the initial distributions}\label{subsec:initial_conditions}

In the following sections, we will apply our methods to an analysis of the combined HERA data  \cite{HERAcombined} on deep inelastic $ep$ scattering. Those data determine the behavior of $F_2^{\gamma p}(x,Q^2)$  very well for $x\lesssim 0.1$ for a wide range of $Q^2$.  $F_2^{\gamma p}(x,Q^2)$ can therefore be taken as accurately known throughout that region through a global fit to the HERA data.

Our objective is to check the consistency of  LO QCD evolution with the HERA data by starting at an initial $Q_0^2$, and evolving or devolving to the final values of $Q^2$ where we can compare  the evolved $F_2^{\gamma p}(x,Q^2)$ directly to the  experimental results. To do this, we need to determine the initial gluon distribution $G(x,Q_0^2)$, which is not measured directly, and the initial singlet distribution $F_s(x,Q_0^2)$, both over the entire range $(x,1)$,  evolve or devolve the distributions as discussed above, and then convert the resulting $F_s(x,Q^2)$ back to $F_2^{\gamma p}(x,Q^2)$. We will discuss the elements of this procedure in the following subsections.


\subsubsection{Determination of $G_0(x)=G(x,Q_0^2)$} \label{subsubsec:G_0}

The LO evolution equation for $F_2^{\gamma p}(x,Q^2)$, easily constructed from the evolution equations for the individual quark distributions and the relation $F_2^{\gamma p}(x,Q^2)=\sum_ie_i^2x\left(q_i+\bar{q}_i\right)(x,Q^2)$, is
\ba
\frac{4\pi}{\alpha_s(Q^2)}\frac{\partial F_2^{\gamma p}(x,Q^2)}{\partial \ln (Q^2)} &=&  4{F_2^{\gamma p}(x,Q^2)}-\frac{16}{ 3}\int_x^1\frac{\partial F_2^{\gamma p}}{\partial z}(z,Q^2)\ln\left(\frac{z-x}{z}\right)dz \nonumber \\
&&  - \frac{8}{3}x\int_x^1F_2^{\gamma p}(z,Q^2)\left(1+\frac{x}{z}\right)\frac{\,dz}{z^2}
 + \sum_i e_i^2 \int_x^1G(z,Q^2)\left(1-\frac{2x}{z}+\frac{2x^2}{z^2}\right)\frac{dz}{z^2}. \label{F2eqn}
 \ea

We have shown elsewhere  \cite{bdm1,bdm2}  that, assuming that  a LO treatment of the DGLAP evolution of $F_2^{\gamma p}$ is consistent, we can invert \eq{F2eqn} to obtain $G(x,Q^2)$ at any given $x,\,Q^2$  directly from a global fit to $F_2^{\gamma p}(x,Q^2)$ which includes the interval $(x,1)$ and a range of $Q^2$ around the desired value. In particular,
\ba
\!\!\!\! G(x,Q^2)\!\!&=\!\!&\!3\,{ \calF}{
\calF}(x,Q^2)-x\frac{\partial { \calF}{ \calF}(x,Q^2)}{\partial x
}-\!\int_x^1 \! {\calF}{ \calF}(z,Q^2) \left(\frac{x}{
z}\right)^{3/2}\left\{\frac{6}{ \sqrt 7} \sin \left[\frac {\sqrt
7}{2} \ln\frac{z}{x}\right]\!\!+2\cos\left[\frac{\sqrt
7}{2}\ln{\frac{z}{ x}}\right]\! \right\}\frac{dz}{ z}, \label{Gofx}
\ea
where ${\cal F}{\cal F}(x,Q^2)$ is the function
\ba
{\cal F}{\cal F}(x,Q^2) &=& \left(\sum_i e_i^2\right)^{-1}\left[\frac{4\pi}{\alpha_s(Q^2)}\frac{\partial F_2^{\gamma p}(x,Q^2)}{\partial \ln (Q^2)} -  4{F_2^p(x,Q^2)}\right. \nonumber \\
\label{FF}
&& \left. +\frac{16}{ 3}\int_x^1\frac{\partial F_2^{\gamma p}}{\partial z}(z,Q^2)\ln\left(\frac{z-x}{z}\right)dz+ \frac{8}{3}x\int_x^1F_2^{\gamma p}(z,Q^2)\left(1+\frac{x}{z}\right)\frac{\,dz}{z^2} \right]
\ea
obtained by combining all the  $F_2^{\gamma p}$-dependent terms in \eq{F2eqn} and dividing the result by $\sum_ie_i^2$.

Since ${\cal F}{\cal F}$ is determined by $F_2^{\gamma p}$,  \eq{Gofx}  determines $G$ directly from experiment   provided the assumption of LO evolution is valid. We have found that the result for $G(x,Q^2)$ at small $x$ is fairly insensitive to the behavior of $F_2^{\gamma p}(x,Q^2)$ at large $x$, so $G(x,Q^2)$ is determined at small $x$ primarily by the HERA data. However, to get precise results, we need a global fit to $F_2^{\gamma p}$ that extends to $x=1$. We will discuss that extension below.

If LO evolution is consistent with the HERA data, the distribution $G(x,Q_0^2)$ determined by \eq{Gofx} should satisfy the gluon evolution equation. We observed very early in our analysis that this condition was not satisfied. In particular, the derivative $\partial G(x,Q^2)/\partial\ln{Q^2}$ was not equal to the sum of terms in the gluon evolution equation that involve weighted integrals of $G$ and $F_s$. While this indicated that the assumption of LO evolution was not consistent, the strength of this conclusion was limited by  the somewhat-limited accuracy with which the derivative of $G$ could be determined. We have therefore adopted the alternative approach that we pursue here, and limit our consistency tests to the evolution of $F_2^{\gamma p}$, where direct comparisons with the HERA data are possible.


\subsubsection{Determination of the singlet distribution $F_s(x,Q^2)$} \label{subsubsec:Fs0}

 In the LO CTEQ6L \cite{CTEQ6L} and MSTW2008LO \cite{MSTW1} analyses which we will use for comparisons, the singlet quark distribution function $F_s(x,Q^2)$ was  determined  through a simultaneous fit to all the quark distributions and the gluon distribution. Those analyses used  earlier variations of the HERA data \cite{H1,ZEUS1,ZEUS2} in combination with other data on deep inelastic electron and neutrino scattering that are concentrated at higher $x$. Because of apparent incompatibilities among various data sets discussed in \cite{CTEQ6L,MSTW1}, and the high accuracy of the combined HERA data at small $x$, we will adopt instead a hybrid approach in which we write $F_s(x,Q^2)$ in terms of $F_2^{\gamma p}(x,Q^2)$ and relatively small nonsinglet quark distributions. We will then take $F_2^{\gamma p}(x,Q^2)$ from a global fit to the combined HERA data, and will use the nonsinglet contributions obtained in the  CTEQ6L and MSTW2008LO analyses  to construct $F_s(x,Q^2)$.  Those analyses differ in their  treatments of $\alpha_s$ in NLO and LO, respectively.

Introducing the  nonsinglet quark distributions \cite{Ellis}
\ba
\label{V_i}
V_i &=& x( q_i-\bar{q}_i),\quad i=1,2,3, \\
\label{T3}
T_3 &=&  x( u+\bar{u}-d-\bar{d}), \\
\label{T8}
T_8 &=&  x( u+\bar{u}+d+\bar{d}-2s-2\bar{s}), \\
\label{T15}
T_{15} &=&  x( u+\bar{u}+d+\bar{d}+s+\bar{s}-3c-3\bar{c}), \\
\label{T24}
T_{24} &=&  x( u+\bar{u}+d+\bar{d}+s+\bar{s}+c+\bar{c}-4b-4\bar{b}),
\ea
we can write $F_s(x,Q^2)$ for different numbers $n_f$ of active quarks as
\ba
\label{Fs_nf=3}
F_s(x,Q^2) &=& \frac{9}{2}F_2^{\gamma p}(x,Q^2)-\frac{3}{4}T_3(x,Q^2)-\frac{1}{4}T_8(x,Q^2), \quad n_f=3, \\
\label{Fs_nf=4}
F_s(x,Q^2) &=&  \frac{18}{5}F_2^{\gamma p}(x,Q^2) -\frac{3}{5}T_3(x,Q^2)-\frac{1}{5}T_8(x,Q^2)+\frac{1}{5}T_{15}(x,Q^2),  \quad n_f=4,  \\
\label{Fs_nf=5}
F_s(x,Q^2) &=& \frac{45}{11}F_2^{\gamma p}(x,Q^2) -\frac{15}{22}T_3(x,Q^2)-\frac{5}{22}T_8(x,Q^2)+\frac{5}{22}T_{15}(x,Q^2)-\frac{3}{22}T_{24}(x,Q^2) , \quad n_f=5.
\ea

Our procedure is now the following. We start with our global fit to $F_2^{\gamma p}(x,Q^2)$ and pick an initial value of $Q^2$ in a region where $F_2^{\gamma p}$ is well determined, here $Q_0^2=4.5$ GeV$^2$, a value between the charm and bottom thresholds. We start by using the nonsinglet distributions $T_3$, $T_8$, and $T_{15}$ from the  CTEQ6L (or MSTW2008LO) fit to the older HERA and high-$x$ data to get an initial result for  the singlet distribution $F_s(x,Q_0^2)$ from $F_2^{\gamma p}(x,Q^2)$ using \eq{Fs_nf=4}. We also determine $G(x,Q_0^2)$ from $F_2^{\gamma p}(x,Q_0^2)$ using \eq{Gofx}.

We next devolve $F_s(x,Q^2)$  to the charm quark threshold at $Q^2=M_c^2$. The $c$ and $\bar{c}$ distributions should vanish at $Q^2=M_c^2$, with $T_{15}(x,M_c^2)=F_s(x,M_c^2)$ for $n_f=3$. However, because we have started with somewhat different data on $F_2^{\gamma p}(x,Q^2)$ than used in earlier analyses, this threshold condition will not be satisfied exactly.  We therefore use the continuity of $F_s(x,Q^2)$ at the $n_f=3$, $n_f=4$ transition, set $T_{15}(x,M_c^2)$ equal to the devolved $F_s(x,M_c^2)$ for $n_f=4$, and evolve $T_{15}$ back to $Q_0^2$ using the LO nonsinglet procedure discussed in \cite{bdhmNLO} to obtain a modified $T_{15}(x,Q_0^2)$. This is used to get a modified $F_s(x,Q_0^2)$ from $F_2^{\gamma p}(x,Q_0^2)$, and the process is repeated if necessary until the result for $T_{15}$ does not change significantly. The changes in $F_s$ introduced by this procedure are small except near the charm threshold. The $F_s(x,Q_0^2)$ obtained from $F_2^{\gamma p}(x,Q_0^2)$ using the modified $T_{15}$ and the CTEQ6L (or MSTW2008LO) distributions $T_3(x,Q_0^2)$ and $T_8(x,Q_0^2)$ gives the initial singlet distribution for use in our subsequent calculations.

The situation with respect to $T_{24}(x,Q^2)$ is simpler. This distribution comes in at the $b\bar{b}$ threshold, where $T_{24}(M_b^2)=F_s(x,M_b^2)$ for $n_f=4$. We therefore determine the initial distribution $T_{24}(x,M_b^2)$ by evolving $F_s(x,Q^2)$ from $Q_0^2$ to $M_b^2$, and its extension to higher $Q^2$, by evolving $T_{24}$ from $M_b^2$ to $Q^2$ using the results of \cite{bdhmNLO} restricted to LO for nonsinglet evolution.

Finally, the evolved or devolved $F_2^{\gamma p}(x,Q^2)$ is determined from evolved or devolved $F_s(x,Q^2)$ for a given $n_f$ using the appropriate one of Eqs.\  (\ref{Fs_nf=3})-(\ref{Fs_nf=5}). The corresponding quark distributions can be obtained from $F_s(x,Q^2)$ and the (modified) nonsinglet distributions, as discussed later.


\subsection{A global fit to the combined HERA data for $F_2^{\gamma p}(x,Q^2)$}\label{subsec:HERA_fit}


The constructions above depend on our having a global fit to the $x$ and $Q^2$ dependence of the structure function $F_2^{\gamma p}(x,Q^2)$. Berger, Block and Tan \cite{bbt2} showed that ZEUS data from HERA \cite{ZEUS1,ZEUS2} could be parametrized accurately as a function of $x$ and $Q^2$ for $x\leq 0.1$ by an expression of the form
\be
 F_2^p(x,Q^2)=(1-x)\left[\frac{F_P}{ 1-x_P}+ A\ln
\frac{x_P(1-x)}{x(1-x_P)}+B\ln^2\frac{x_P(1-x)}{x(1-x_P)}\right].
\label{Fofx}
\ee
We will use the same parametrization for the complete HERA data sets as combined in \cite{HERAcombined}.

In the expression in \eq{Fofx},  $x_P$ specifies the location in $x$ of an approximate fixed point observed in the data where curves of $F_2^{\gamma p}(x,Q^2)$ for different $Q^2$ cross. At that point, $\partial F_2^{\gamma p}(x_P,Q^2)/\partial \ln Q^2\approx 0$ for all $Q^2$;
$F_P=F_2^{\gamma p}(x_P,Q^2)$ is the common value of $F_2^{\gamma
p}$.  The $Q^2$ dependence of $F_2^{\gamma p}(x,Q^2)$ is  given in
those fits by
\ba
    A(Q^2)&=&a_0+a_1\ln Q^2 +a_2\ln^2 Q^2, \quad
    B(Q^2)=b_0+b_1\,\ln Q^2 +b_2\,\ln^2 Q^2.  \label{AB}
\ea

We used this parametrization to fit the combined HERA  data  for $Q^2\gtrsim 1$ GeV$^2$. These data included 34 different $Q^2$ values with $x\leq 0.11$, specifically,  $Q^2=$
0.85, 1.2, 1.5, 2.0, 2.7, 3.5, 4.5, 6.5, 8.5, 10, 12, 15, 18, 22,
27, 35, 45, 60, 70, 90, 120, 150, 200, 250, 300, 400, 500, 650,
800, 1000, 1200, 1500, 2000, and 3000 GeV$^2$. The scaling
point value $x_P=0.11$ was taken to be fixed.

The data set has a total of 356 datum points. The use of  the
sieve algorithm to sift the data to eliminate outliers as described  in \cite{sieve}   eliminated 14 points whose contribution to the $\chi^2$ of the fit was 125.0, roughly a quarter of the total.    The values of the 7 fit parameters, along with their statistical
errors, are given in Table \ref{table:results}.  The fit using the sieve algorithm gives a minimum with $\chi^2_{\rm min}= 352.8$. This must be corrected by the sieve factor ${\cal R}=1.109$ to account for the change in normalization of the $\chi^2$ function \cite{sieve}. This gives a corrected value ${\cal R}\times\chi^2_{\rm min}=391.4$, so a corrected $\chi^2$ per degree of freedom of 1.17, a reasonable result for this much data.

For the 296 points with $Q^2\geq 2.7$ GeV$^2$ that we will consider later,  the fit is excellent, with  $\chi^2=295$. For comparison,  the CTEQ6L \cite{CTEQ6L} and MSTW2008LO \cite{MSTW1} fits, made using the separate H1 \cite{H1} and ZEUS \cite{ZEUS1,ZEUS2} data rather than the combined results, give $\chi^2$ of 3339 and 1329, respectively, with uncorrected values of $\chi^2$/d.o.f. of 11.3 and 4.5.

\begin{figure} [ht]
\begin{center}
\mbox{\epsfig{file=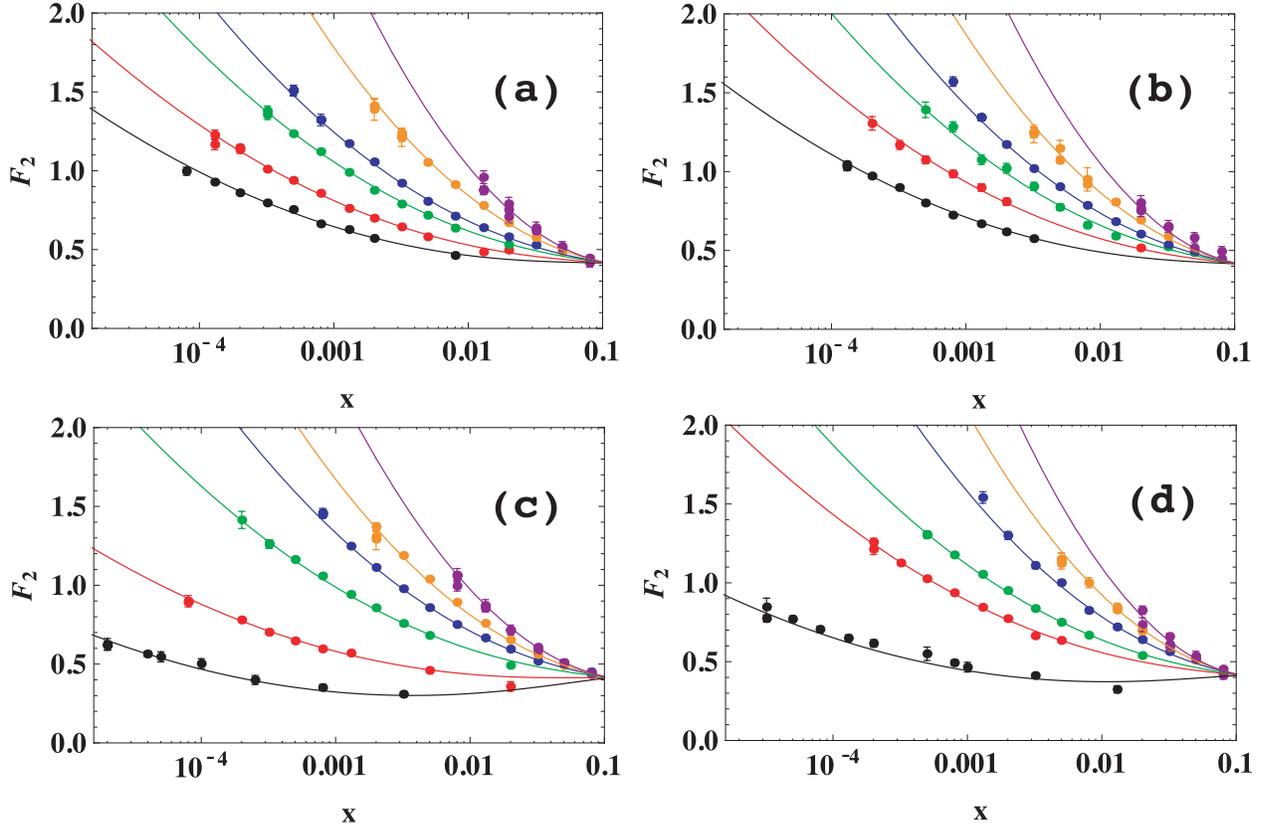,width=6.5in }}
\end{center}
\caption[]{Comparison of our fit to the proton structure function $F_2^{\gamma p}(x,Q^2)$  with the combined HERA data \cite{HERAcombined}, plotted as functions of the Bjorken variable $x$, with $Q^2$ increasing from the bottom to the top curves in each panel: (a) $Q^2=3.5,\,6.5,\,15,\,27,\,120,\,650$ GeV$^2$; (b) $Q^2=4.5,\,10,\,22,\,45,\,150,\,800$ GeV$^2$; (c) $Q^2=0.85,\,2.7,\,12,\,35,\,90,\,400$ GeV$^2$; (d) $Q^2=1.5,\,8.5,\,18,\,70,\,250,\,1200$ GeV$^2$.  The fixed point in the fit was taken as $x_P=0.11$.} \label{fig:F2_HERA_fit}
\end{figure}

Curves of the fitted $F_2^{\gamma p}(x,Q^2)$ plotted  as a function $x$ are compared with the data in Fig. \ref{fig:F2_HERA_fit} for 24 values of $Q^2$.  The quality of the fit is evident.

We emphasize that our fitting procedure is quite different from that used in other analyses.  Our fit is directly to $F_2^{\gamma p}(x,Q^2)$, and its adequacy can be tested at that level. An investigation of possible alternative models with more parameters gave essentially equivalent results in the experimental region. We use the model in \eq{Fofx} and \eq{AB} because of its simplicity and its reasonable behavior for small $x$ \cite{bbt2}, and more importantly, for its excellent $\chi^2$ fit with a minimum number of parameters. With this approach, our fit to the HERA data  is independent of any assumptions about QCD evolution, and will allow us later to obtain a direct test of the validity  of purely LO evolution.  In contrast, the usual methods, such as those in \cite{CTEQ6L,MSTW1,CTEQ6.1}, start by  assuming the validity of QCD evolution to some order in the strong coupling $\alpha_s$, calculate $F_2^{\gamma p}$ from a complete set of  parton distributions evolved from some initial $Q_0^2$, and then attempt to fit the data by adjusting the (many) parameters in the initial parton distributions.


\begin{table}[ht]                   
%
\begin{center}
\def\arraystretch{1.2}            
     \caption{\label{fitted}\protect\small Results of a 7-parameter fit to the
HERA combined data for $F_2^{\gamma p}(x,Q^2)$ for $0.85 \le
Q^2\le 3000$ GeV$^2$.\label{table:results}}
\begin{tabular}[b]{|l||c||}
    \multicolumn{1}{l}{Parameters}&\multicolumn{1}{c} {Values}\\
\hline
      $a_0$&$-8.471\times 10^{-2}\pm 2.62\times 10^{-3}$ \\
      $a_1$&$\phantom{-}4.190\times 10^{-2}\pm 1.56\times 10^{-3}$\\
      $a_2$&$-3.976\times 10^{-3}\pm 2.13\times 10^{-4}$\\
\hline
    $b_0$ &$\phantom{-}1.292\times 10^{-2}\pm 3.62\times 10^{-4}$\\
      $b_1$&$\phantom{-}2.473\times 10^{-4}\pm 2.46\times 10^{-4}$\\
      $b_2$&$\phantom{-}1.642\times 10^{-3}\pm 5.52\times 10^{-5}$ \\
\hline
$F_P$&$0.413\pm0.003$\\
    \cline{1-2}
        \hline
    \hline
    $\chi^2_{\rm min}$&352.8\\
    ${\cal R}\times\chi^2_{\rm min}$&391.4\\
    d.o.f.&335\\
\hline
	${\cal R}\times\chi^2_{\rm min}$/d.o.f.&1.17\\
\hline
\end{tabular}
\end{center}
\end{table}
\def\arraystretch{1}  


\subsection{Extension of the fit to high $x$}\label{subsec:FGhi}

Our fit to the data on  $F_2^{\gamma p}(x,Q^2)$ is so far
restricted to the region  $x\le x_P$; we have not attempted to fit
the DIS data for $x>x_P$ from other experiments. Since the
expressions for the evolved $F_s(x,Q^2)$ and $G(x,Q^2)$ in terms of their initial distributions at $Q_0^2$ given in Eqs.\ (\ref{F}) and (\ref{G}), and that for $G$ in terms of $F_2^{\gamma p}$ given in \eq{Gofx}, involve integrals that extend to $x=1$,  we need also to extend the parametrization of $F_2^{\gamma p}(x,Q^2)$  to $x=1$.  We will again use the results of earlier analyses, this time less directly, in making the extension.

We have found that the CTEQ6L and MSTW2008LO versions of $F_2^{\gamma p}(x,Q_0^2)$ for $Q_0^2=4.5$ GeV$^2$ are well approximated at large $x$ by expressions of the form
\be
\label{Fhi}
F_2^{\gamma p}(x,Q^2) = F_0\left(\frac{x}{
x_0}\right)^{\mu(Q^2)}\left(\frac{1-x}{1-x_0}\right)^n\frac{1+ax+bx^2}{1+ax_0+bx_0^2}, \quad1\geq  x\geq x_0.
\ee
We will use this form to extend our fit to $F_2^{\gamma p}(x,Q^2)$ to the high-$x$ region,  where the HERA data are restricted to values of $Q^2$ much larger than our chosen $Q_0^2$, and $F_2^{\gamma p}(x,Q_0^2)$ is not well determined.  In making this extension, we must choose the starting $x_0$ sufficiently small that we avoid problems with our lack of precise knowledge of the $x$ and $Q^2$ dependence of $F_2^{\gamma p}(x,Q^2)$ for $x$ near the fixed point in our fit. We have used $x_0=0.03$ in the present calculations. With this choice, the CTEQ6L result for $F_2^{\gamma p}$ is well fitted with $a=6.83$, $b=13.0$, and $n=3.75$ in \eq{Fhi}. For MSTW2008LO, $a=4.83$, $b=13.7$, and $n=3.66$.

We match the expression in \eq{Fhi} in value and slope at  $x_0=0.03$ to the expression in \eq{Fofx} which describes the HERA data by adjusting the parameters $F_0$ and $\mu$, retaining the initial values of $a$, $b$, and $n$. The changes necessary in $\mu$ are fairly small, with increases of 4.6\% and 4.0\% in  magnitude from the CTEQ6L and MSTW2008LO values, respectively. The changes in the normalizations are somewhat larger, 11.8\% and 8.7\%. To a good approximation, the extended distributions in the region $x>0.03$ are simply scalings of the CTEQ6L and MSTW2008LO results for $F_2^{\gamma p}(x,Q_0^2)$, retaining the shapes of those distributions. Our final results at small $x$ are insensitive to the details of these extensions.

The determination of the initial gluon distribution at $Q_0^2$ involves further complications.  As discussed in Sec.\ \ref{subsubsec:G_0},  $G(x,Q_0^2)$ can be determined directly from $F_2^{\gamma p}(x,Q^2)$. It can be shown from Eqs.\ (\ref{Gofx}) and (\ref{FF}) that $G(x,Q_0^2)$ is actually determined mainly by $F_2^{\gamma p}(x,Q_0^2)$ and its derivative $\partial F_2^{\gamma p}(x,Q^2)/\partial \ln{Q^2}$ at $Q_0^2$; the integral terms in \eq{FF} are small. The need to know $\partial F_2^{\gamma p}(x,Q^2)/\partial \ln{Q^2}$ introduces some complication because the fixed point imposed in \eq{Fofx} reflects the observed $Q^2$ dependence of  $F_2^{\gamma p}(x,Q^2)$ for $x$ near $ x_P=0.11$ only qualitatively, and not precisely. The HERA data near that point are restricted to $Q^2>>Q_0^2$, and do not determine  $\partial F_2^{\gamma p}(x,Q^2)/\partial \ln{Q^2}$ in the region $Q^2\approx Q_0^2$ where it is needed.  The derivative at $Q_0^2$ is, in fact,  only determined well by the fit to the HERA data for $x<<x_P$ and  $Q^2\approx Q_0^2$. In particular, the fit to $F_2^{\gamma p}$ and its extension to high $x$ do not give reliable results on its $Q^2$ dependence for $x>>0.03$. As a result, the expression in \eq{Gofx} cannot be used to determine $G$ in that region.

We therefore adopt an approach similar to that used with $F_2^{\gamma p}$. We choose a small value of $x_0$, $x_0=0.03$ where $F_2^{\gamma p}$ and its $Q^2$ dependence are well determined, and determine $G(x,Q_0^2)$  for $x\leq x_0$ from the  fit to $F_2^{\gamma p}(x,Q^2)$ using \eq{Gofx}.  The small uncertainties in the extensions of $F_2^{\gamma p}$ to large $x$ affect only the integral terms in Eqs.\ (\ref{Gofx}) and (\ref{FF}), and do not affect the  result  for $G$ significantly in the region of concern, $x\leq 0.03$.

To extend the result for $G$ to higher $x$, we  fit the shapes of the gluon distributions $G(x,Q_0^2)$ given by  CTEQ6L and MSTW2008LO for $x>x_0=0.03$ using the same functional form as in \eq{Fhi}. We use the results to extend $G$  to $x>x_0$ by adjusting the analogs of the parameters $\mu$ and $F_0$ so that the extensions match the  $G$ derived for $x<x_0$  in magnitude and slope at $x=x_0$. The result is a gluon distribution $G(x,Q_0^2)$ that retains the basic shape of the CTEQ6L or MSTW2008LO gluon distribution  for $x>x_0$, merges smoothly into the form derived from $F_2^{\gamma p}$ for $x\leq x_0$, and, in contrast to other analyses, involves no {\em a priori} assumptions about the form of $G$ in the latter region.


\section{Applications to the HERA data on $F_2^{\gamma p}(x,Q^2)$}\label{sec:application_HERA}


In this section, we summarize the results we obtained by applying our methods to an analysis of the HERA data on deep inelastic electron-proton scattering as combined by the H1 and ZEUS experimental groups \cite{HERAcombined}.

We first examine the consistency of our results for $F_2^{\gamma p}(x,Q^2)$, $G(x,Q^2)$ , and the quark distributions with other LO results, represented here by CTEQ6L and MSTW2008LO.  We find qualitative, but not quantitative agreement, with our evolved $F_2^{\gamma p}(x,Q^2)$ agreeing much better with the combined HERA data, and our $G(x,Q^2)$ generally increasing much less rapidly at small $x$ than the distributions found elsewhere. These changes will affect the results of cross section and other calculations performed using LO quark and gluon distributions.

We then turn to a central question, the consistency of a LO treatment of the QCD evolution, and  examine the consistency of the structure function $F_2^{\gamma p}(x,Q^2)$ determined by LO evolution with the  HERA data. We  conclude on the basis of {\em systematic,} $Q^2$ -{\em dependent discrepancies} that LO evolution cannot give an adequate description of those data. At least NLO corrections are needed. We emphasize that this conclusion is independent of any calculation of the NLO corrections, and follows directly from the $Q^2$ dependence of the data.


\subsection{Basic results and comparisons with other analyses}\label{subsec:comparisons}

\subsubsection{Starting distributions and sum-rule tests}\label{subsubsec:starting_distributions}

Our results are based on the smooth global fit to the measured $F_2^{\gamma p}(x,Q^2)$ discussed in Sec.\ \ref{subsec:HERA_fit}. The fit was very good, as seen in Fig.\ \ref{fig:F2_HERA_fit}, and determined our starting distributions at $Q_0^2=4.5$ GeV$^2$, a value chosen in the region of dense data where the $x$ and $Q^2$ dependence of $F_2^{\gamma p}$ are tightly constrained.

\begin{figure} [ht]
\begin{center}
\mbox{\epsfig{file=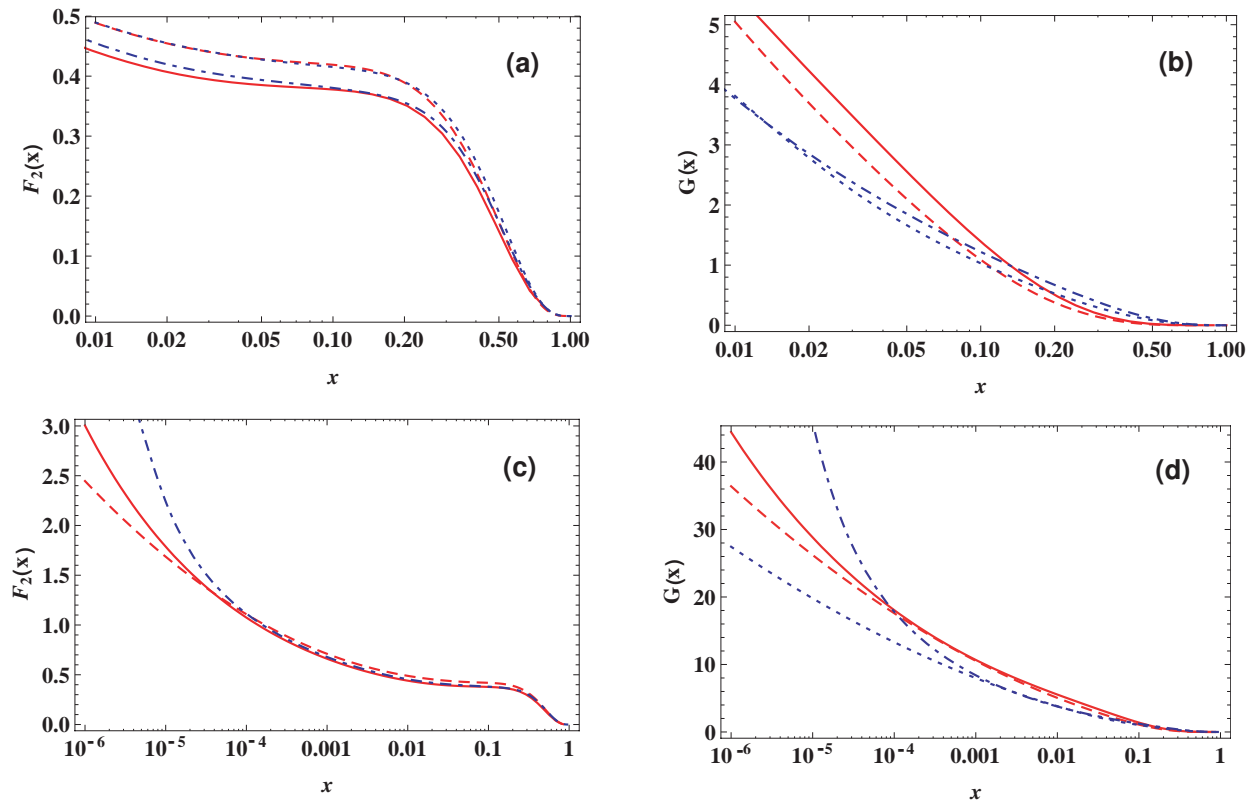,width=6.5in }}
\end{center}
\caption[]{Comparison of our starting distributions $F_2^{\gamma p}(x,Q_0^2)$ and $G(x,Q_0^2)$ with those of  CTEQ6L and MSTW2008LO at $Q_0^2=4.5$ GeV$^2$. (a) $F_2^{\gamma p}$ from our fit to the HERA data, extended to $x>0.03$ using the method described in the text based on the  CTEQ-like (red dashed lines) or  MSTW-like (blue dotted lines) shape of $F_2^{\gamma p}$ at larger $x$. The original CTEQ6L (red solid lines) and MSTW2008LO (blue dot-dashed lines) versions of $F_2^{\gamma p}$ are shown for comparison. (b)  The $G(x,Q_0^2)$  derived from  $F_2^{\gamma p}$ using the condition that $F_2^{\gamma p}$ satisfy its DGLAP evolution equation in LO, using the NLO (red dashed lines)  or LO (blue short dashed) versions of $\alpha_s$, compared to the corresponding CTEQ6L (red solid lines) and MSTW2008LO (blue dot-dashed lines) distributions. (c) Extension of (a) to small $x$. The CTEQ-like and MSTW-like shapes for $F_2^{\gamma p}$ for $x>0.03$ cannot be distinguished on the scale of the figure, and only the former is shown. (d) Extension of (b) to small $x$.} \label{fig:F20_G0_comparisons}
\end{figure}

$F_2^{\gamma p}(x,Q_0^2)$ is fixed by the fit. We  determined the  initial $G(x,Q_0^2)$ directly from the fit to $F_ 2^{\gamma p}(x,Q^2)$ using \eq{Gofx} and the extensions to high $x$ discussed in Sec.\ \ref{subsec:FGhi}. The uncertainty in our derived $G$ at small $x$ is determined mainly by $\partial F_2^{\gamma p}(x,Q^2)/\partial\ln{Q^2}$, and is quite small \cite{bdm1}.
We compare these initial distributions with those that resulted from the CTEQ6L and MSTW2008LO analyses  in Fig.\ \ref{fig:F20_G0_comparisons}.  There are clearly significant differences in the magnitudes and $x$ dependence of distributions among the sets.

We note first in Fig.\ \ref{fig:F20_G0_comparisons}(a) that the difference between the extensions of $F_2^{\gamma p}(x,Q_0^2)$ for $x>0.03$ we obtain for CTEQ6L-like and MSTW2008LO-like shapes is very small. These differences lead  to negligible effects in the evolution of $F_2^{\gamma p}$ and $G$ at small $x$. The differences evident between our curve for $F_2^{\gamma p}$ and those shown for CTEQ and MSTW in Fig.\ \ref{fig:F20_G0_comparisons}(a)  result from their failure to fit this quantity accurately, presumably attributable in part to their use of the older H1 and ZEUS versions of the data.

The differences in our curves for $G$ in  Fig.\ \ref{fig:F20_G0_comparisons}(b) from those of the CTEQ6L and MSTW2008LO analyses result from the difference between their $F_2^{\gamma p}$ and ours. The marked difference between the curves shown for our CTEQ-like and MSTW-like gluon distributions results from the different treatments of $\alpha_s$ used by the two groups, which we follow here.   CTEQ6L treats $\alpha_s$ to NLO, with
\ba
\label{alpha_s}
\alpha_s(Q^2)
&=&\frac{4 \pi}{\beta_{0}\ln(Q^2/\Lambda^2)}\left[1-\frac{2\beta_1}{\beta_0^2}\frac{\ln[\ln(Q^2/\Lambda^2)]}{\ln(Q^2/\Lambda^2)}\right], \\
\beta_0(n_f)&=& 11-{\frac{2}{3}}n_f, \qquad \beta_1(n_f)=51-\frac{19}{3}n_f.
\label{alphasNLO}
\ea
The value of $\alpha_s$ is fixed to the measured value at the $Z$-boson mass, $\alpha_s(M_Z^2)=0.118$ for $n_f=5$, and the value of $\Lambda(n_f)$ is then adjusted at the $b$ and $c$ thresholds where $n_f$ decreases by 1 to assure continuity.

MSTW2008LO, in contrast, uses only the first, LO, term in \eq{alphasNLO}  for presumed consistency in a LO analysis, and treats the value of $\alpha_s$ at $Q^2=1$ GeV$^2$ as a parameter in their fitting procedure. This leads to a value $\alpha_s(M_Z^2)=0.139$.
The two versions of $\alpha_s$ do not agree well, with the MSTW2008LO version being considerable larger at all $Q^2$. We note that the $Q^2$ dependence of $\alpha_s$ is actually well determined by experiment  \cite{average_alpha_s}, with the results well described by the NLO expression  \cite{pdg2010_alpha_s} for $\alpha_s$ fixed to $\alpha_s(M_Z^2)$. Since $F_2^{\gamma p}$ is also known, the CTEQ-like determination of $G(x,Q_0^2)$ is based entirely on measured quantities, with the assumption that a LO  analysis of the evolution is adequate. Our MSTW-like approach uses the MSTW2008LO version of $\alpha_s$, but at the expense of poor agreement with the measured $Q^2$ dependence of $\alpha_s$.

 Figures\ \ref{fig:F20_G0_comparisons}(c) and (d) show the extensions of the curves in 2(a) and 2(b) to small $x$.
We emphasize that with the assumption that the LO evolution equation for $F_2^{\gamma p}$ is satisfied, a necessary condition for a consistent LO analysis, our initial  gluon distribution $G$ at $Q_0^2=4.5$ GeV$^2$ follows directly from our global fit to the $x$ and $Q^2$ dependence of the HERA data on $F_2^{\gamma p}(x,Q^2)$ and its extension to large $x$. In this sense,  $F_2^{\gamma p}(x,Q_0^2)$, $G(x,Q_0^2)$, and up to small corrections, $F_s(x,Q_0^2)$ are all determined by experiment for $x\gtrsim  10^{-4}$ where there are substantial HERA data, and determined to lesser accuracy down to $x\sim 10^{-5}$ where the data at presumably perturbative values of $Q^2$ run out. It is not necessary to determine these quantities indirectly through initial  parametrizations of the complete set of quark distributions and $G$, with the many parameters determined only in a fit to the data.

We conclude that the strong divergences  of $F_2^{\gamma p}$ and $G$ evident in the MSTW2008LO curves in Figs.\ \ref{fig:F20_G0_comparisons}(c) and (d) are not realistic in a LO analysis. The lesser differences between the CTEQ6L results and ours in Figs.\ \ref{fig:F20_G0_comparisons}(c) and (d) are mainly in the region $x<10^{-5}$ where some extrapolation from the data is necessary, so it is less definitive.


Following the procedures discussed in  Sec.\ \ref{subsubsec:Fs0}, we used the  fit to $F_2^{\gamma p}(x,Q^2)$ and  the results for the nonsinglet quark distributions $V_i$, $T_3$, $T_8$ and the initial $T_{15}$ given by CTEQ6L or MSTW2008LO, to determine the corresponding  LO result for $F_s(x,Q_0^2)$.

As a test of our  procedures, we  evaluated the QCD momentum sum rule, which should give
\be
\label{momentum_sum}
\int_0^1dx\left[F_s(x,Q_0^2)+G(x,Q_0^2)\right]=1.
\ee
We find that it is satisfied to $\sim  0.1$\% (1.2\%) at $Q^2=4.5$ GeV$^2$ for the $F_s$ and $G$ derived from the extended fit to $F_2^{\gamma p}$  using the nonsinglet distributions from CTEQ6L (MSTW2008LO) and the method of Sec.\ \ref{subsubsec:Fs0}.   Because of the structure of the splitting functions, the sum rule for the evolved distributions is automatically satisfied to similar accuracy at all $Q^2$.

This result may seem startling: the CTEQ6L and MSTW2008LO results for $F_s(x,Q_0^2)$ and $G(x,Q_0^2)$ also satisfy the sum rule, used in those fits as a constraint, but our global fit to the combined HERA data at $Q_0^2$ lies considerably above the $F_2^{\gamma p}(x,Q_0^2)$ calculated from their quark distributions as  seen in Fig.\ \ref{fig:F20_G0_comparisons}, with similar differences in $F_s$. However, the gluon distribution $G(x,Q_0^2)$, calculated from the requirement that $F_2^{\gamma p}$ satisfy its LO DGLAP evolution equation exactly, is smaller than the $G$ obtained in other analyses in the region of $x$ that contributes significantly to the sum rule, as seen in Fig.\ \ref{fig:F20_G0_comparisons}.

The two effects compensate for each other numerically. The contributions to the momentum sum rule from $F_s$ and $G$ at $Q_0^2=4.5$ GeV$^2$ are 0.550 (0.622) and  0.452 (0.377) for the CTEQ6L (CTEQ6L-like) distributions, with the calculated sum rule equal to  1.002 (0.999).  The results for the MSTW2008LO (MSTW-like) distributions are similar, with contributions to the sum rule from $F_s$ and $G$  of 0.565 (0.637) and 0.434 (0.373)  at $Q_0^2=4.5$ GeV$^2$, for total of 0.999 (1.010). We have {\em  not} used the sum rule as a constraint, as is done in other analyses. Its satisfaction follows from the data and our determination of $G$ in terms of $F_2^{\gamma p}$. We conclude that our extensions of $F_2^{\gamma p}$ and $G$ to the large-$x$ region cause no problems.

The quark number sum rules
\be
\label{number_sums}
\int_0^1 dx \left(u-\bar{u}\right)(x,Q_0^2)=2, \qquad  \int_0^1 dx\left( d-\bar{d}\right)(x,Q_0^2)=1,
\ee
are different. Because we set the nonsinglet  distributions $u-\bar{u}$ and $d-\bar{d}$  equal to the corresponding CTEQ6L or MSTW2008LO distributions and do not change them in our hybrid analysis, the quark number sum rules
 are satisfied automatically to the extent that they were satisfied by the  CTEQ6L and MSTW  distributions, namely to $\sim0.4$\% ($\sim 0.5$\%).  The changes introduced in the separate $u$ and $\bar{u}$, and $d$ and $\bar{d}$ distributions by the changes in $F_2^{\gamma p}$ and $F_s$, are confined to the singlet combinations $u+\bar{u}$ and $d+\bar{d}$, and cancel in the differences $u-\bar{u}$ and $d-\bar{d}$.

 The corresponding sum rules for $s-\bar{s}$, $c-\bar{c}$, and $b-\bar{b}$ give zero in the CTEQ6L-based analysis since those quarks are produced only in pairs through gluon splitting. For the MSTW2008LO-based input, $s\not= \bar{s}$ initially. The very small difference is not changed in our analysis because we keep the nonsinglet distributions fixed, and the strange-quark sum rule remains constant at $\approx 0.0028$. The $c$, $\bar{c}$ and $b$, $\bar{b}$ quarks are produced only in pairs, and the quark sum rules give zero.

\subsubsection{Leading-order gluon and quark distributions}

\begin{figure} [ht]
\begin{center}
\mbox{\epsfig{file=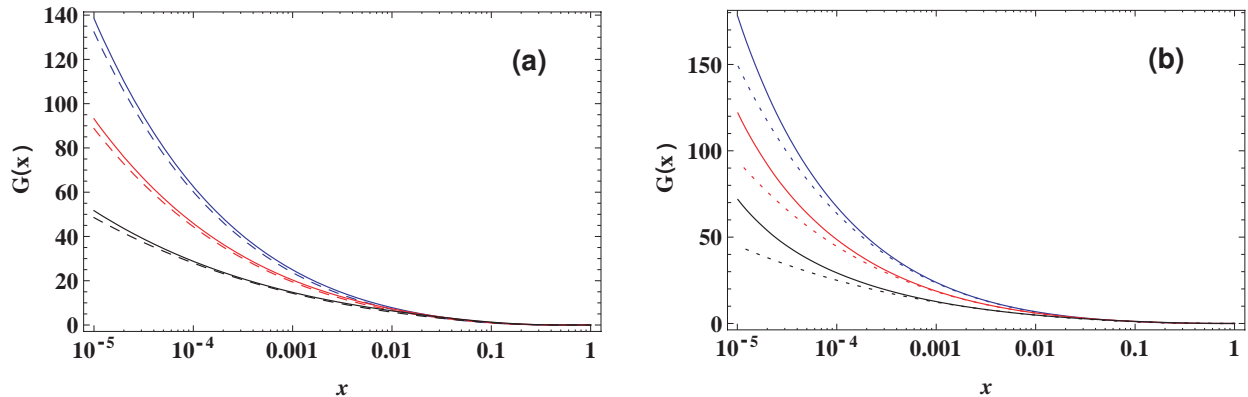,width=6.5in }}
\end{center}
\caption[]{Comparison of our evolved gluon distributions $G(x,Q^2)$ with the CTEQ6L and MSTW2008LO distributions. (a) Our $G$  (dashed curves) at $Q^2=10$ (black), 35 (red), and 120 (blue) GeV$^2$ , bottom to top, compared to the CTEQ6L $G$ (solid curves). (b) Our $G$ (dotted curves) at $Q^2=10$ (black), 35 (red), and 120 (blue) GeV$^2$ , bottom to top, compared to the MSTW2008LO $G$ (solid curves).} \label{fig:G_comparison}
\end{figure}

 We evolved the starting distributions for $F_s$ and $G$ from $Q_0^2$ to lower and higher values of $Q^2$ using the Laplace transform methods sketched in Sec.\ \ref{subsec:solutions}, using the numerical techniques discussed  in the Appendix to \cite{inverseLaplace2}. We compare the evolved gluon distributions $G(x,Q^2)$ to those of CTEQ6L \cite{CTEQ6L} and MSTW2008LO \cite{MSTW1} in Fig.\ \ref{fig:G_comparison}.


It is evident from the figure that  our gluon distributions are somewhat smaller than those of CTEQ6L and MSTW2008LO, quite significantly so for the latter at small values of $x$ where MSTW uses a strongly power-law divergent parametrization with their initial $G(x,Q_0^2)$.  Our CTEQ- and MSTW- based results also differ significantly, the result of the differing initial distributions seen in Fig.\ \ref{fig:F20_G0_comparisons} and the different treatments of $\alpha_s$ as NLO and LO respectively.

It is straightforward to combine our results for $F_s(x,Q^2)$ with the original nonsinglet distributions $V_i$, $T_3$, $T_8$, and the modified $T_{15}$ and $T_{24}$,  Eqs.\ (\ref{V_i})-(\ref{T24}) to obtain the quark distributions that lead to these results.  The results differ from the individual quark distributions given by CTEQ6L and MSTW2008LO because of the changes in the HERA data, and, more importantly, because of  our treatment of the starting distributions for the evolution of $F_s(x,Q^2)$ and $G(x,Q^2)$.

\begin{figure} [ht]
\begin{center}
\mbox{\epsfig{file=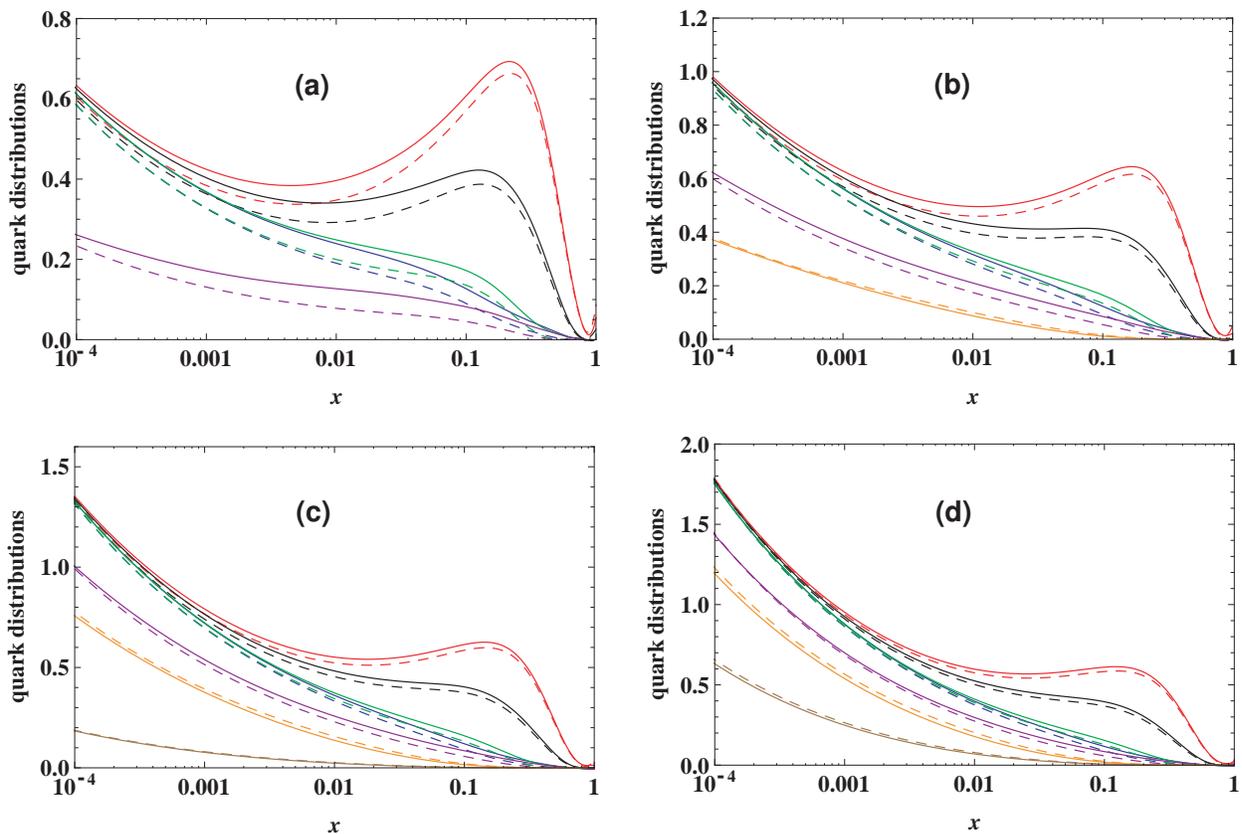,width=6.5in }}
\end{center}
\caption[]{Plots of the quark distributions obtained by our method using the nonsinglet distributions from CTEQ6L \cite{CTEQ6L}, shown for: (a), $Q^2=M_c^2=1.69$ GeV$^2$; (b), $Q^2= 10$ GeV$^2$; (c),  $Q^2= 35$ GeV$^2$; and (d),  $Q^2= 120$ GeV$^2$. The solid lines give our distributions. The dashed lines are the CTEQ6L distributions.  The curves show $xq(x,Q^2)$ for, top to bottom in each panel at $x\approx 0.1$, the $u$ (red), $\bar{u}$ (blue), $d$ (black), $\bar{d}$ (green), $s=\bar{s}$ (purple), $c=\bar{c}$ (orange), and $b=\bar{b}$ (brown) quarks} \label{BDHM_quarks_CTEQ6L}
\end{figure}

\begin{figure} [ht]
\begin{center}
\mbox{\epsfig{file=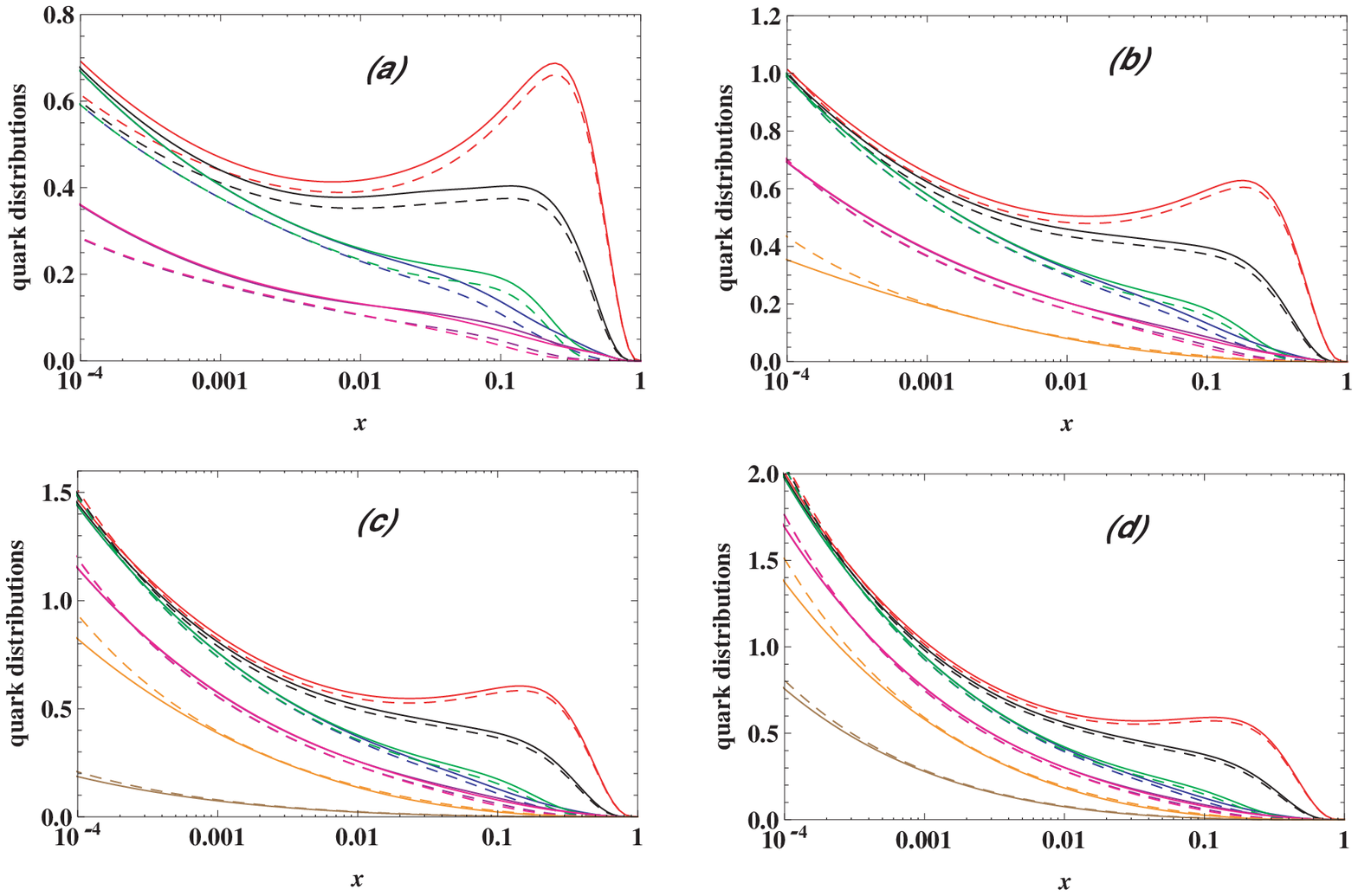,width=6.5in }}
\end{center}
\caption[]{Plots of the quark distributions obtained by our method using the nonsinglet distributions from MSTW2008LO \cite{MSTW1}, shown for: (a), $Q^2=M_c^2=1.96$ GeV$^2$; (b), $Q^2= 10$ GeV$^2$; (c),  $Q^2= 35$ GeV$^2$; and (d),  $Q^2= 120$ GeV$^2$. The solid lines give our distributions. The dashed lines are the MSTW2008LO distributions.  The curves show $xq(x,Q^2)$ for, top to bottom in each panel at $x\approx 0.1$, the $u$ (red), $\bar{u}$ (blue), $d$ (black), $\bar{d}$ (green), $s$ (purple), $\bar{s}$ (magenta), $c=\bar{c}$ (orange), and $b=\bar{b}$ (brown) quarks.} \label{BDHM_quarks_MSTW}
\end{figure}

Our results for the quark distributions are shown in Figs.\ \ref{BDHM_quarks_CTEQ6L} and \ref{BDHM_quarks_MSTW} for the treatments based on the CTEQ6L and MSTW2008LO nonsinglet terms, respectively.  The differences from the input distributions are not large in the region of the HERA data, but some changes are evident at higher values of $x$, and, especially for MSTW, at very small $x$.  We attribute the differences to the   parametrizations of the quark and gluon distributions used by those authors, which have a strong power-law dependence on $1/x$ at small $x$, with the many parameters adjusted to fit the data used.

Our method is based instead on our overall fit to $F_2^{\gamma p}(x,Q^2)$, and the information that can be derived from it. It uses the earlier nonsinglet distributions only in calculating small terms involved in  the transitions between $F_2^{\gamma p}$ and $F_s$. The results on the fit shown in Fig.\ \ref{fig:F2_HERA_fit} suggest  that its $x$ and $Q^2$ dependence are well determined for  $Q^2$ of a few GeV$^2$ for $x>10^{-5}$. This allows the reliable derivation of the starting distributions  needed in the solution of the LO evolution equations in that region. In that sense, our results are as reliable as allowed by the assumption of strict LO evolution. They do not depend on choices of parametrizations for initial quark and gluon distributions. The results shown in Figs.\  \ref{fig:G_comparison}, \ref{BDHM_quarks_CTEQ6L}, and \ref{BDHM_quarks_MSTW} follow.


\subsection{Check of the consistency of LO DGLAP evolution with the HERA data}\label{subsec:LO_consistency}


As a final application of our methods, we turn to the question of the consistency of LO evolution with experiment. We  show that the structure functions $F_2^{\gamma p}(x,Q^2)$ obtained by LO evolution from the initial distributions at $Q_0^2$ determined by the HERA data  are not consistent with the data at higher and lower values of $Q^2$.  A consistent analysis must therefore include higher-order terms in $\alpha_s$ in the evolution equations, and distributions evolved out of the experimental region using the LO DGLAP equations cannot be used with confidence.

\begin{figure}[h,t,b]
\begin{center}
\mbox{\epsfig{file=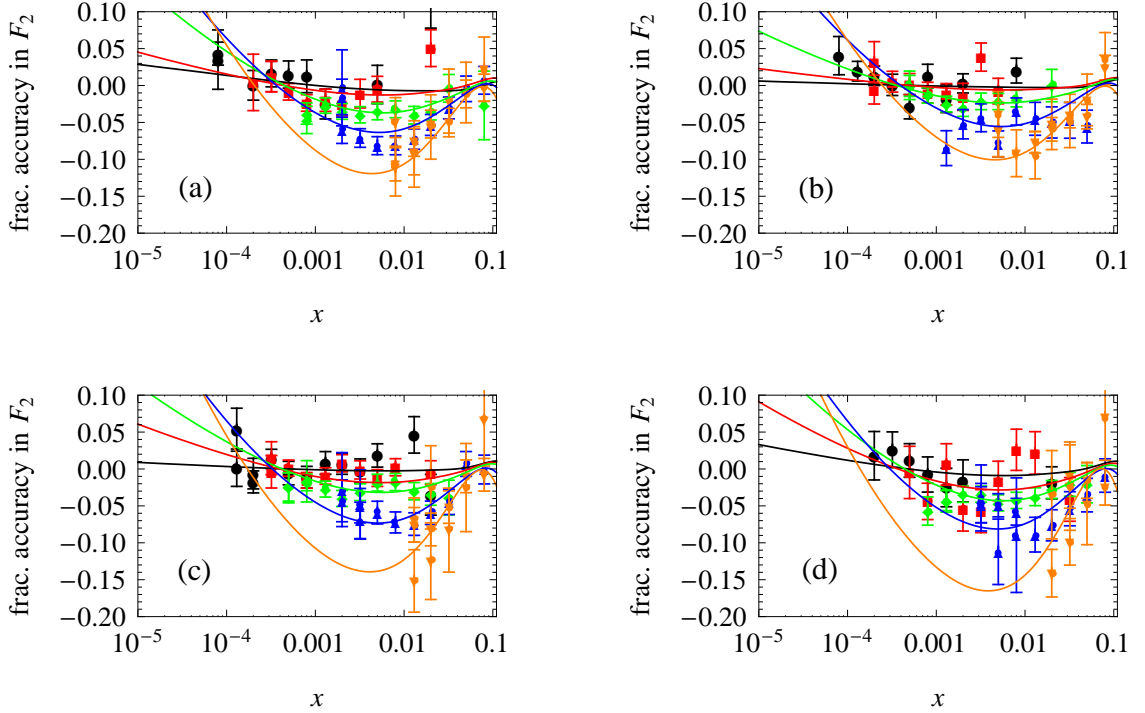
,width=6.5in
}}
\end{center}
\caption[]{Fractional accuracy  $\left(F_{2,{\rm evolved}}^{\gamma p}-F_{2,{\rm HERA}}^{\gamma p}\right)/F_{2,{\rm HERA}}^{\gamma p}$ evolved from $Q_0^2=4.5$ GeV$^2$ relative to our fit to the combined HERA data \cite{HERAcombined}, compared to the same ratio with the data for $F_2^{\gamma p}$ used in the numerator. The initial and final conversions between $F_2^{\gamma p}$ and the singlet distribution $F_s$ are based on the CTEQ6L nonsinglet quark distributions \cite{CTEQ6L}. Results are given for (a) $Q^2= 2.7$ (black dots), 12 (red squares), 35 (green diamonds), 90 (blue triangles), 400 (orange inverted triangles) GeV$^2$; (b)  $Q^2=3.5$ (black dots), 8.5 (red squares), 18 (green diamonds), 70 (blue triangles), 250 (orange inverted triangles) GeV$^2$; (c)  $Q^2=6.5$ (black dots), 15 (red squares), 27 (green diamonds), 120 (blue triangles), 650 (orange inverted triangles) GeV$^2$; and (d)  $Q^2=10$ (black dots), 22 (red squares), 45 (green diamonds), 150 (blue triangles), 1200 (orange inverted triangles) GeV$^2$.
}\label{fig:F2_frac_accuracy_CTEQ6L}
\end{figure}

\begin{figure}[h,t,b]
\begin{center}
\mbox{\epsfig{file=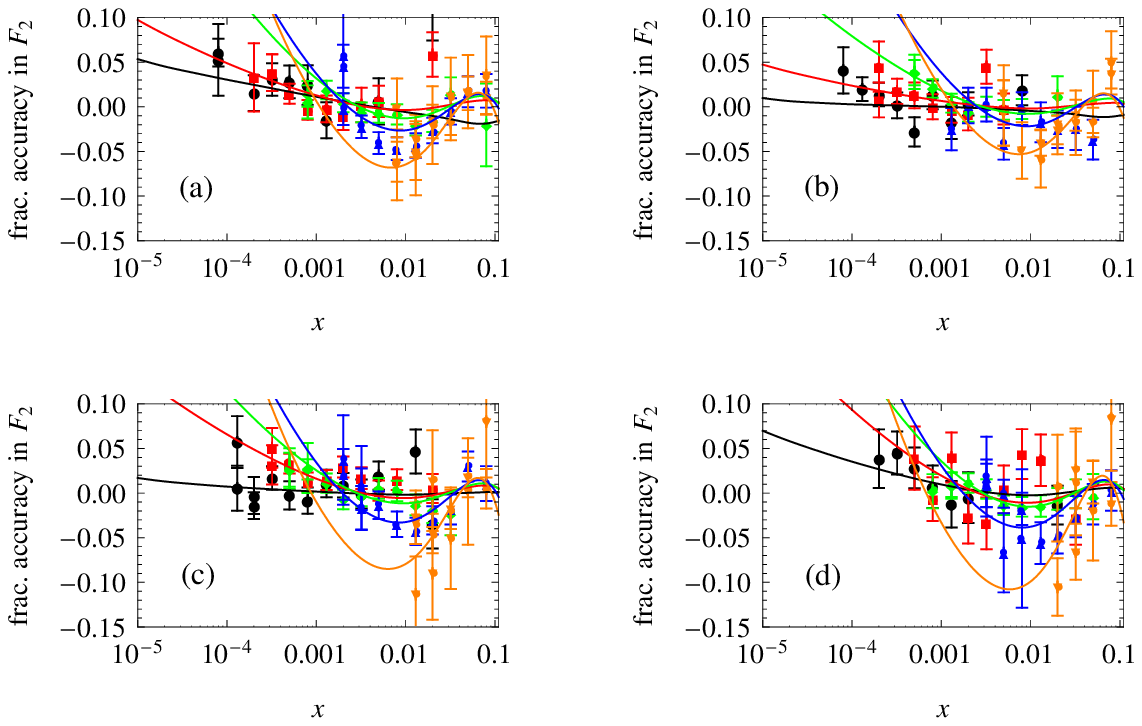
,width=6.5in
}}
\end{center}
\caption[]{Fractional accuracy $\left(F_{2,{\rm evolved}}^{\gamma p}-F_{2,{\rm HERA}}^{\gamma p}\right)/F_{2,{\rm HERA}}^{\gamma p}$ of the structure function  $F_2^{\gamma p}(x,Q^2)$ evolved from $Q_0^2=4.5$ GeV$^2$ relative to our fit to the combined HERA data \cite{HERAcombined}, compared to the same ratio with the data for $F_2^{\gamma p}$ used in the numerator. The initial and final conversions between $F_2^{\gamma p}$ and the singlet distribution $F_s$ are based on the MSTW2008LO nonsinglet quark distributions \cite{MSTW1}. Results are given for (a) $Q^2= 2.7$ (black dots), 12 (red squares), 35 (green diamonds), 90 (blue triangles), 400 (orange inverted triangles) GeV$^2$; (b)  $Q^2=3.5$ (black dots), 8.5 (red squares), 18 (green diamonds), 70 (blue triangles), 250 (orange inverted triangles) GeV$^2$; (c)  $Q^2=6.5$ (black dots), 15 (red squares), 27 (green diamonds), 120 (blue triangles), 650 (orange inverted triangles) GeV$^2$; and (d)  $Q^2=10$ (black dots), 22 (red squares), 45 (green diamonds), 150 (blue triangles), 1200 (orange inverted triangles) GeV$^2$.
}\label{fig:F2_frac_accuracy_MSTW}
\end{figure}

We plot the ratios $\left(F_{2,{\rm evolved}}^{\gamma p}-F_{2,{\rm HERA}}^{\gamma p}\right)/F_{2,{\rm HERA}}^{\gamma p}$ for 20 values of $Q^2$ where there are data in  Figs. \ref{fig:F2_frac_accuracy_CTEQ6L} and \ref{fig:F2_frac_accuracy_MSTW}.  Here $F_{2,\rm evolved}^{\gamma p}$ is the distribution evolved (or devolved) from $Q_0^2=4.5$ GeV$^2$, and $F_{2,\rm HERA}^{\gamma p}$ is our fit to the HERA data. We also show the ratios with  $F_{2,\rm HERA}^{\gamma p}$ replaced in the numerators by the actual data points.

We can see from the figures that the evolved distributions differ {\em systematically} from the fit and the data,   falling too low for $Q^2>Q_0^2$ for $x$ in the range $\sim 5\times 10^{-4}-10^{-1}$, and too high for $x\lesssim 5\times 10^{-4}$. The discrepancies increase systematically with increasing $Q^2$, span about a 10\% range for $0.001\lesssim x\lesssim 0.01$, and have the same pattern for the analyses based on the CTEQ6L and MSTW2008LO nonsinglet distributions.  The datum points follow the curves, as they should; the problem is not in the fit. The systematic increase of the discrepancies with increasing $Q^2$ indicates that they are the result of incorrect evolution at LO, with the evolved $F_2^{\gamma p}$ not growing sufficiently rapidly with $Q^2$. We conclude that LO DGLAP evolution of $F_2^{\gamma p}$ is inconsistent with the combined HERA data.

The systematic trends are evident quantitatively in Table \ref{chisq_table}. Using the 296 data points in our sample of the combined HERA data for $Q^2\geq 2.7$ GeV$^2$, we find  a $\chi^2$  ($\chi^2$ per degree of freedom) of 295 (0.996) for our fit from Sec.\ \ref{subsec:HERA_fit},   1480 (5.00) for the evolved $F_2^{\gamma p}$ that used  the CTEQ6L nonsinglet terms to convert between $F_2^{\gamma p}$ and $F_s$, and 502 (1.70) for the evolved $F_2^{\gamma p}$ that used the nonsinglet distributions of MSTW2008LO.  Our direct fit to the HERA data is quite good given the large amount of data, with probability $P=0.126$ when $\chi^2$ is corrected for the sieve factor \cite{sieve} ${\cal R}=1.109$. The evolved distributions have essentially zero probabilities of being correct statistically.

The difference in the values of $\chi^2$ for the CTEQ6L- and MSTW2008LO-based treatments of the nonsinglet terms is the result primarily of the different treatments of $\alpha_s$ in the two cases. The NLO treatment in CTEQ6L is fixed to the value of $\alpha_s$ at $M_Z^2$, and agrees well with the measured values of $\alpha_s$ down to  $M_\Upsilon^2$. In contrast, the value of the LO version of $\alpha_s$ at $Q^2=1$ GeV$^2$ is used  in MSTW2008LO as a fitting parameter. The result is an $\alpha_s$ that is larger than the NLO version by about 40\% at $Q^2=1$ GeV$^2$, and 18\% at $M_Z^2$, so it does not agree with the measured values. This results in rather different starting distributions at $Q_0^2=4.5$ GeV$^2$ in the two cases, as seen in Fig.\ \ref{fig:F20_G0_comparisons}, and to more rapid QCD evolution in the case of the MSTW2008LO-based treatment. Although the resulting $\chi^2$ is reduced, the systematic problems with the evolved $F_2^{\gamma p}$ remain, as seen in Fig.\ \ref{fig:F2_frac_accuracy_MSTW}, and the result is still unacceptable statistically.

This failure of LO evolution to give an accurate description of the separate H1 and ZEUS  data has been noted in \cite{CTEQ6.1,MSTW1}, and no doubt elsewhere, in connection with poor values of the $\chi^2$ for $F_2^{\gamma p}$ obtained in LO in those analyses, and the improvements afforded by a NLO treatment of the parton distributions.  The systematic nature of the problem is somewhat obscured there by the way initial conditions are imposed through many-parameter descriptions of the complete set of parton distributions, and the subsequent adjustment of those parameters to minimize the $\chi^2$ of the fit.


\begin{table}[ht]
\begin{center}
   \caption{The $\chi^2$ of  the $F_2^{\gamma p}$ from our fit to the combined HERA data, and of the evolved  $F_2^{\gamma p}$ obtained  by LO evolution from $Q_0^2=4.5$ GeV$^2$. The starting distribution $G(x,Q_0^2)$ was derived from the fit to $F_2^{\gamma p}$. The initial $F_s(x,Q_0^2)$ was obtained from  $F_2^{\gamma p}$ using nonsinglet corrections from the CTEQ6L \cite{CTEQ6L} and MSTW2008LO \cite{MSTW1} analyses. The last lines give sums of all the rows above them except for the starting value $Q_0^2=4.5$ GeV$^2$.}
   \label{chisq_table}
  \begin{tabular}{|c|c|c|c|c|}
  \hline

  $Q^2$ (in GeV $^2$) & No. of datum points & \ $\chi^2$ (our fit)\  &{ \begin{tabular}{c} $\chi^2_{\rm evolved}$, CTEQ \\
  corrections \end{tabular}}  &\begin{tabular}{c} $\chi^2_{\rm evolved}$, MSTW \\ corrections\end{tabular}  \\
  \hline

2.7 &  9 & 10.0 & 15.4 & 28.0  \\

3.5 & 9 & 11.1 & 11.6 & 11.5 \\

4.5  & 9  & 6.1 &   6.1   &  6.1   \\

6.5 & 13 & 14.2 & 13.6 & 14.3 \\

8.5 & 9 &  7.6 & 7.6 & 10.7 \\

10   & 7  &  2.4 &  3.8 & 7.1 \\

12  &  10 & 11.5  & 15.1  & 19.5  \\

15   & 10 & 10.6 &   5.0 &  28.5 \\

18  & 9  & 2.74  &  25.0 &  15.9 \\

22 & 9 & 12.4 & 14.0 & 10.4 \\

27   & 12  & 9.1 &  52.9  & 15.4  \\

35   & 11 &  8.8  &  81.2 & 11.0  \\

45 & 11 & 8.0 & 96.9 & 7.3 \\

60  & 10  & 17.2  & 158.9  & 19.4  \\

70  & 9  & 13.7  & 68.7  & 12.6  \\

90  & 11 & 13.0 &  175.9 &  49.1  \\

120 & 12 &  6.8 & 102.0  &  25.1  \\

150  & 12  & 15.9  & 71.0  & 16.5  \\

200  & 14  & 21.0  & 114.6  & 33.5  \\

250 & 14 & 15.8 &  86.6  & 25.6   \\

300  &  15 &  18.9 &  83.5 &  24.4 \\

400  &  14 &  18.7 &  76.0 & 21.6  \\

500 & 11 & 5.8 & 37.7 & 18.7 \\

650  & 12  & 10.4  & 57.5  & 21.4  \\

800  &  9 &  11.0 &  40.8 & 18.8  \\

1000  & 9  & 6.1  & 13.3  & 5.9  \\

1200 & 9 &  10.0 &  33.1  & 18.2  \\

1500  &  6 &  5.8 &  9.9 & 5.7  \\

2000  & 5  & 0.33  & 0.26  &  1.1 \\

3000  & 5  & 6.3  & 7.5  &  4.7 \\

  \hline
 \vspace*{0.5ex}
Sum (without $Q^2=4.5$)  & 296 & 295.2 & 1480 & 502 \\
$\chi^2$/d.o.f. && 1.003 & 5.00 & 1.70 \\
  \hline

  \end{tabular}
\end{center}
\end{table}



\section{Summary and conclusions}\label{sec:conclusions}


In the present paper, we have applied recently developed  methods  based on Laplace transforms to a  LO analysis of the HERA data on deep inelastic $ep$ scattering as combined by the H1 and ZEUS experimental groups  \cite{HERAcombined}.  We have used a hybrid method, in which we convert the measured  structure function $F_2^{\gamma p}(x,Q^2)$ to the singlet distribution $F_s(x,Q^2)$ which enters the evolution equations, taking  the small contributions of nonsinglet quark distributions to this conversion  from other analyses, and extending the fit to the HERA data for $x<0.1$ to $x=1$ using the shape of $F_2^{\gamma p}$ determined in those analyses. Here we used the results of the CTEQ6L \cite{CTEQ6L} and MSTW2008LO \cite{MSTW1} analyses, which used the older H1 \cite{H1} and ZEUS \cite{ZEUS1,ZEUS2} data along with data from other experiments, mostly at  higher values of $x$ than the HERA data. This procedure determines the starting distribution $F_s(x,Q_0^2)$ at the starting point $Q_0^2=4.5$ GeV$^2$ chosen for the DGLAP evolution.

 As shown earlier \cite{bdm1,bdm2}, the necessary starting distribution $G_0(x)\equiv G_0(x,Q_0^2)$ for the coupled evolution of $F_s$ and $G$  can be obtained in LO directly from a global fit to the structure function $F_2^{\gamma p}(x,Q^2)$  by requiring that the LO evolution equation for $F_2^{\gamma p}(x,Q^2)$ be satisfied for $Q^2=Q_0^2$. Both $F_{20}(x) \equiv F_2^{\gamma p}(x,Q_0^2)$ and $G_0(x)$ are therefore determined  directly by experiment through our fit to the HERA data for $x<0.1$ and its extension to higher $x$, without the need for a solution of the complete set of coupled parton evolution equations or any assumptions about the functional form of $G$.  Our results at small $x$ are insensitive to the details of the extensions.

 We picked  a starting value  $Q_0^2=4. 5$ GeV$^2$ for the evolution which is well within the region of dense data.
We then solved the LO evolution equations using very fast and accurate methods discussed elsewhere \cite{inverseLaplace1,inverseLaplace2}, and combined the evolved $F_s$ with the evolved nonsinglet distributions of CTEQ6L and MSTW2008LO to obtain a new set of  quark distributions. These differ from the quark distributions obtained in those analyses because of our use of the combined HERA data rather than the original H1 and ZEUS results, and our different determination of the starting distributions in $F_s$ and $G$ for the evolution. The differences in the quark distributions are significant in some regions. Our gluon distributions differ markedly from those of MSTW2008LO at small $x$ as seen in Figs.\  \ref{fig:F20_G0_comparisons} and \ref{fig:G_comparison}.

Finally, we compared the evolved structure function $F_2^{\gamma p}(x,Q^2)$ to the HERA data as a test of the consistency of LO DGLAP evolution. The initial distributions of $F_s$ and $G$ at $Q_0^2$ were determined by  $F_2^{\gamma p}$ up to the small nonsinglet corrections,  and were  consistent with LO evolution by construction.
We concluded  that LO evolution is actually {\em not} consistent with those data on the basis of {\em systematic trends} evident  in the evolved distributions. This conclusion does not depend on the explicit calculation of NLO effects. It is supported by a $\chi^2$ analysis, but in contrast to other approaches, {\em we could not attempt to reduce the} $\chi^2$ by adjusting the shapes of the initial distributions: {\em we had no arbitrary parameters to adjust.}

In the Appendix, we give an equally accurate, though approximate, method which works directly with the exact DGLAP LO evolution equation for $F_2^{\gamma p}$ coupled to an approximate evolution equation for $G$. This approach is independent of the nonsinglet distributions, and its implementation uses only the experimental results as extended above.  The results of the analysis are the same: LO evolution of $F_2^{\gamma p}$ is inconsistent with the HERA data.


\begin{acknowledgments}

The authors would like to  thank the Aspen Center for Physics, where this work was supported in part by NSF Grant No.\ 1066293, for its hospitality during the time parts of this work were done. M. M. B. would like to thank Professor Arkady Vainstein of the University of Minnesota for many valuable discussions.   P. H. would like to thank Towson University Fisher College of Science and Mathematics for travel support.  D.W.M. receives support from DOE Grant No. DE-FG02-04ER41308.

\end{acknowledgments}


\appendix

\section{Approximate coupled evolution equations for $F_2^{\gamma p}(x,Q^2)$ and $G(x,Q^2)$}\label{subsec:F2_G_equations}

In this Appendix, we point out that we can obtain a direct test of the adequacy of LO evolution using evolution equations coupling $F_2^{\gamma p}(x,Q^2)$ and $G(x,Q^2)$. In particular, we use the exact LO evolution equation for $F_2^{\gamma p}$, and an approximate version of the evolution equation for $G$ in which $F_s$ is replaced by a multiple of $F_2^{\gamma p}$.

The advantage of this approach is that it deals directly with the experimentally accessible function $F_2^{\gamma p}(x,Q^2)$,  and gives  a direct test of the adequacy of LO evolution with no input beyond a global fit to $F_2^{\gamma p}$. It does not require direct knowledge of the nonsinglet quark distributions, but correspondingly does not provide individual quark distributions unless $V_i$, $T_3$, $T_8$, $T_{15}$, and $T_{24}$ are known. If these are to be used, the method developed in Sec.\ \ref{sec:preliminaries} is to be preferred.

The results on the evolution of $F_2^{\gamma p}(x,Q^2)$ from its initial distribution at $Q_0^2=4.5$ GeV$^2$ obtained by this method differ insignificantly from those obtained with the method in the body of the paper, with fractional differences small on the scale of the differences of the evolved $F_2^{\gamma p}$ from the data shown in Figs.\ \ref{fig:F2_frac_accuracy_CTEQ6L} and \ref{fig:F2_frac_accuracy_MSTW}. We conclude again that the assumption LO evolution is not consistent with the HERA data.

We obtain our evolution equations for $F_2^{\gamma p}$ and $G$ as follows. The exact LO evolution equation for $F_2^{\gamma p}(x,Q^2)$ is given in \eq{F2eqn}. This equation couples $F_2^{\gamma p}$ to the gluon distribution $G$. The exact evolution equation for $G$ couples $G$ instead to the singlet quark distribution $F_s(x,Q^2)=\sum_ix\left(q_i+\bar{q}_i\right)(x,Q^2)$, and not to $F_2^{\gamma p}=\sum_ie_i^2x\left(q_i+\bar{q}_i\right)(x,Q^2)$.
$F_s(x,Q^2)$ is not determined directly by experiment.  However, we note that the nonsinglet contributions in the transition from $F_2^{\gamma p}$ to $F_s$  given in Eqs.\ (\ref{Fs_nf=3})-(\ref{Fs_nf=5}) are very small, and will simply replace $F_s(x,Q^2)$ in the usual evolution equation for $G(x,Q^2)$ by the leading, $F_2^{\gamma p}$-dependent terms  in Eqs.\ (\ref{Fs_nf=3})-(\ref{Fs_nf=5}), $F_s(x,Q^2)\approx a(n_f) )F_2^{\gamma p}(x,Q^2)$ with  $a(n_f)=18/5$ for $M_c^2<Q^2<M_b^2$, and 45/11 for $M_b^2<Q^2<M_t^2$ \footnote{ This is the result obtained in the approximation that all sea-quark distributions are taken as the same, and that the sea quarks dominate in $F_s$ and $F_2^{\gamma p}$ outside the valence region.}. These relations are actually only expected to hold for $Q^2$ well above  thresholds, where the new quarks can be taken as fully excited; we will use them as stated.

We use the resulting approximate evolution equation for $G$  with the exact LO evolution equation for $F_2^{\gamma p}$ in \eq{F2eqn},  and solve for $F_2^{\gamma p}$ and $G$ using the  methods developed earlier \cite{bdm1,bdhmLO2,bdhmNLO}.  The accuracy of the method  is evident from Table\ \ref{approx_accuracy_F2_G_method}, where  we compare the results for the evolved $F_2^{\gamma p}$ obtained using the approximate method with those obtained using the exact evolution equations for $F_s$ and $G$ and the CTEQ6L nonsinglet corrections in the $F_2^{\gamma p}$, $F_s$ transition as described in Sec.\ \ref{subsubsec:Fs0}. The accuracy is similar for the MSTW2008LO-based nonsinglet corrections.


\begin{table}[h]
\begin{center}
     \caption{Fractional differences $\Delta_{F}$ and $\Delta_{G}$ (in \%)  between the $F_2^{\gamma p}$ and $G$ distributions obtained using the ``exact'' transformation between $F_s$ and $F_2^{\gamma p}$  described in Sec.\ \ref{subsubsec:Fs0}, and those obtained using the approximate method based on the exact evolution equation for $F_2^{\gamma p}$, and an approximate gluon evolution equation with $F_s$ replaced by a multiple of $F_2^{\gamma p}$ as described in this Appendix. The same starting distributions for $F_2^{\gamma p}$ and $G$ at $Q_0^2=4.5$ GeV$^2$ were used in both cases. In the ``exact'' method, we used the  nonsinglet terms from CTEQ6L to convert between $F_s$ and $F_2^{\gamma p}$. The last column shows the percentage rms differences between the distributions from the two methods for $10^{-6} \le x < 0.5$. Results obtained using the nonsinglet terms from MSTW2008LO  are very similar.}
  \label{approx_accuracy_F2_G_method}
   \begin{tabular}{|c|c|c|c|c|c|c||c|}

   \hline
   & \multicolumn{6}{c||} {\raisebox{-1ex}{$\Delta_F=1-F_2^{\rm approx}/F_2^{\rm exact}$ (\%)}} & {\raisebox{-1ex}{$\Delta_{F, {\rm rms}}$ (\%)}}\\[2ex] \cline{2-7}

      $Q^2$ (in GeV$^2$) & $x=10^{-6}$  & $x=10^{-5}$ & $x=10^{-4}$  & $x=10^{-3}$ & $x=10^{-2}$ & $x=10^{-1}$ & $10^{-6} \le x < 0.5$  \\  \hline

           1.69 & 0.4  & 0.3  & 0.1  & 0.0  & $-$0.2  & $-$0.2 & 0.2  \\

           3.5 & 0.2  & 0.0  & 0.0  & $-$0.1  & $-$0.1  & $-$0.1 & 0.1  \\

           10  & 0.2 & 0.1  & 0.0  & 0.0  & $-$0.1  & $-$0.1 & 0.1  \\

           22  & 0.2  & 0.1  & 0.1  & 0.0  & $-$0.2  & $-$0.1 & 0.1  \\

           27  & 0.2 & 0.1  & 0.1  & 0.1  & $-$0.2  & $-$0.2 & 0.1  \\

           90 & 0.1 & 0.0  & $-$0.1   & $-$0.2  & $-$0.4  & $-$0.3 & 0.2  \\

            250  & $-$0.1 & $-$0.2  & $-$0.3  & $-$0.5  & $-$0.7  & $-$0.4 & 0.4 \\

            1200 & $-$0.4  & $-$0.5  & $-$0.6  & $-$0.9  & $-$1.1  & $-$0.6 & 0.8 \\

            \hline

             & \multicolumn{6}{c|} {\raisebox{-2ex}{$\Delta_G=1-G^{\rm approx}/G^{\rm exact}$ (\%)}} & \raisebox{-2ex}{{$\Delta_{G, {\rm rms}}$ (\%)}} \\[3ex]  \cline{2-7}

      $Q^2$ (in GeV$^2$) & $x=10^{-6}$  & $x=10^{-5}$ & $x=10^{-4}$  & $x=10^{-3}$ & $x=10^{-2}$ & $x=10^{-1}$ & $10^{-6} \le x < 0.5$ \\  \hline

           1.69  & $-$10.6 & $-$3.6  & $-$1.7  & $-$0.8  & 0.1  & 1.8 & 4.6 \\

           3.5  & $-$0.3 & $-$0.2  & $-$0.2  & $-$0.1  & 0.0  & 0.4 & 0.7 \\

           10  & 0.4 & 0.3  & 0.2  & 0.1  & $-$0.2  & $-$1.2 & 1.8  \\

           22  & 0.4 & 0.3  & 0.2  & 0.0  & $-$0.5  & $-$2.3 & 3.3  \\

           27 & 0.3  & 0.2  & 0.1  & $-$0.2  & $-$0.7  & $-$2.8 & 3.8\\

           90 & $-$0.3  & $-$0.4  & $-$0.6  & $-$1.0  & $-$2.0  & $-$5.8 & 6.4 \\

            250 & $-$0.5  & $-$0.7  & $-$1.0  & $-$1.6  & $-$2.9  & $-$7.9 & 8.2  \\

            1200 & $-$0.9  & $-$1.2  & $-$1.6  & $-$2.3  & $-$4.1  & $-$10.6 & 10.4 \\

   \hline
        \end{tabular}
        \end{center}
\end{table}


The approximation of replacing $F_s$ by a multiple of $F_2^{\gamma p}$ is only good to about 5-7\% at $Q^2=5$ GeV$^2$, a value above the $c$-quark threshold but below the $b$-quark threshold, and also at 100 GeV$^2$, well above the $b$ threshold, so  the effect of these errors on the final $F_2^{\gamma p}$ is clearly greatly reduced by the nature of the evolution.
We can understand this qualitatively as follows: the evolution of $G$ at small $x$  is driven mainly by $G$ itself, which is accurately known at the initial  $Q_0^2$ from the condition that the measured $F_2^{\gamma p}(x,Q_0^2)$ satisfy its evolution equation. The final errors in $G$ are therefore small at small $x$, and their effect on $F_2^{\gamma p}$ is further suppressed by the contributions from $F_2^{\gamma p}$ itself to its evolution.  In addition,  $G$ is small at large $x$, and errors in the approximate $G$ in that region have little effect on the  final $F_2^{\gamma p}$.  Overall, the limited accuracy of the approximate $F_s$  has only a small effect on the evolved $G$, and as a result, even less effect on the exact evolution of $F_2^{\gamma p}$ from its known initial distribution.

We have described the methods we use to solve the coupled evolution equations for $\hat{F}_s(v,Q^2)$ and $\hat{G}(v,Q^2)$ in detail elsewhere \cite{bdhmLO2,bdhmNLO}. We use the same methods here to solve the coupled equations for $\hat{F}_2(v,Q^2)$ and $\hat{G}(v,Q^2)$, so we only point out the changes. We begin with Eqs.\ (\ref{f(s,tau)}) and (\ref{g(s,tau)}) which express the Laplace transforms $f_2(s,\tau)$ and $g(s,\tau)$ of the distribution functions in terms of their initial distributions and, here, a set of new kernels $k_{ij}(s,\tau)\rightarrow k^{(2)}_{ij}(s,\tau)$.

The kernels have the same form as those given in  \cite{bdhmLO2,bdhmNLO}, but with the coefficient functions $\Phi$, $\Theta$ that appear there replaced by functions
 $\Phi^{(2)}$ and $\Theta^{(2)}$,
\ba
\Phi_f^{(2)} (s)&=&4 -\frac{8}{3}\left(\frac{1}{ s+1}+\frac{1}{ s+2}+2\left(\psi(s+1)+\gamma_E\right)\right), \label{Phif} \\
\Theta_f^{(2)} (s )&=& \sum_ie_i^2\left(\frac{1}{ s+1}-\frac{2}{s+2}+\frac{2}{ s+3} \right), \label{Thetaf} \\
\Phi_g^{(2)} (s)&=& \frac{33-2n_f }{ 3} +12\left(\frac{1}{ s}-\frac{ 2}{ s+1}+\frac{1}{ s+2}-\frac{1 }{ s+3}-\psi(s+1)-\gamma_E\right) ,\label{Phig} \\
\Theta_g^{(2)} (s)&=& \frac{8}{ 3} a(n_f) \left(\frac{2}{ s}-\frac{2}{ s+1}+\frac{1}{ s+2}  \right). \label{Thetag}
\ea

These functions differ from the corresponding functions in the case of $F_s,\ G$  in the coefficients in the $\Theta$'s, hence the introduction of the superscripts 2 to distinguish the two cases. The kernels $k_{ij}^{(2)}$ have the same formal structure as the original $k_{ij}$, and the final solutions  are obtained as described in Sec.\ \ref{subsec:solutions} using the very fast and accurate algorithms for calculating inverse Laplace transforms introduced in \cite{inverseLaplace1,inverseLaplace2}. The methods needed in practice are discussed in the Appendix of \cite{bdhmLO2}. The results are essentially the same as those presented in Sec.\ \ref{subsec:LO_consistency}, and we draw the same conclusions as there.


\bibliography{G_F2_evolution}


\end{document}